\documentclass[structabstract]{aa}  
\usepackage{graphicx,natbib}
\usepackage{lscape}
\usepackage{longtable}
\usepackage{color}
\usepackage{txfonts}

\newcommand{\mi}{\relax \ifmmode {\mu{\mbox m}}\else $\mu$m\fi}
\newcommand{\sii}{\relax \ifmmode {\mbox S\,{\scshape ii}}\else S\,{\scshape ii}\fi}
\newcommand{\siii}{\relax \ifmmode {\mbox S\,{\textsc {iii}}}\else S\,{\scshape iii}\fi}
\newcommand{\siv}{\relax \ifmmode {\mbox S\,{\textsc {iv}}}\else S\,{\scshape iv}\fi}
\newcommand{\nii}{\relax \ifmmode {\mbox N\,{\scshape ii}}\else N\,{\scshape ii}\fi}
\newcommand{\neii}{\relax \ifmmode {\mbox Ne\,{\textsc {ii}}}\else Ne\,{\scshape ii}\fi}
\newcommand{\neiii}{\relax \ifmmode {\mbox Ne\,{\textsc {iii}}}\else Ne\,{\scshape iii}\fi}
\newcommand{\oiii}{\relax \ifmmode {\mbox O\,{\scshape iii}}\else O\,{\scshape iii}\fi}
\newcommand{\oii}{\relax \ifmmode {\mbox O\,{\scshape ii}}\else O\,{\scshape ii}\fi}
\newcommand{\oi}{\relax \ifmmode {\mbox O\,{\scshape i}}\else O\,{\scshape i}\fi}
\newcommand{\ha}{\relax \ifmmode {\mbox H}\alpha\else H$\alpha$\fi}
\newcommand{\hep}{\relax \ifmmode {\mbox H}\epsilon\else H$\epsilon$\fi}
\newcommand{\hdel}{\relax \ifmmode {\mbox H}\delta\else H$\delta$\fi}
\newcommand{\hgam}{\relax \ifmmode {\mbox H}\gamma\else H$\gamma$\fi}

\newcommand{\pa}{\relax \ifmmode {\mbox Pa}\alpha\else Pa$\alpha$\fi}
\newcommand{\hb}{\relax \ifmmode {\mbox H}\beta\else H$\beta$\fi}
\newcommand{\rdostres}{\relax \ifmmode {\,\mbox{R}}_{\rm 23}\else \,\mbox{R}$_{\rm 23}$\fi}
\newcommand{\ergs}{\relax \ifmmode {\,\mbox{erg\,s}}^{-1}\else \,\mbox{erg\,s}$^{-1}$\fi}
\newcommand{\me}{\relax \ifmmode {\,}^{-1}\else \,$^{-1}$\fi}

\newcommand{\msun}{\relax \ifmmode {\,\mbox{M}}_{\odot}\else \,\mbox{M}$_{\odot}$\fi}

\newcommand{\cmtres}{\relax \ifmmode {\,\mbox{cm}}^{-3}\else \,\mbox{cm}$^{-3}$\fi}
\newcommand{\cmdos}{\relax \ifmmode {\,\mbox{cm}}^{-2}\else \,\mbox{cm}$^{-2}$\fi}
\newcommand{\cmseis}{\relax \ifmmode {\,\mbox{cm}}^{-6}\else \,\mbox{cm}$^{-6}$\fi}
\newcommand{\hi}{\relax \ifmmode {\mbox H\,{\scshape i}}\else H\,{\scshape i}\fi}

\newcommand{\hii}{\relax \ifmmode {\mbox H\,{\textsc ii}}\else H\,{\scshape ii}\fi}

\usepackage[colorlinks]{hyperref}
\begin{document}
   \title{Evolution of the grain size distribution in galactic discs}
   \author{
     M.\,Rela\~{n}o\inst{1,2}  \and
     U.\,Lisenfeld\inst{1,2} \and
     K.-C., Hou\inst{3} \and
     I. \,De Looze\inst{4,5} \and
    J.M. V\'{\i}lchez\inst{6} \and 
    R. C. Kennicutt\inst{7}
   }
   \institute{
     Dept. F\'{i}sica Te\'orica y del Cosmos, Universidad de Granada, Spain -- \email{mrelano@ugr.es}
    \and
      Instituto Universitario Carlos I de F\'isica Te\'orica y Computacional, Universidad de Granada, 18071, Granada, Spain
     \and
     Physics Department, Ben-Gurion University of the Negev, Be'er-Sheva 84105, Israel 
     \and 
     Department of Physics and Astronomy, University College London, Gower Street, London WC1E 6BT, UK
     \and 
     Sterrenkundig Observatorium, Universiteit Gent, Krijgslaan 281 S9, B-9000 Gent, Belgium
     \and
     Instituto de Astrof\'{\i}sica de Andaluc\'{\i}a - CSIC, Glorieta de la Astronom\'{\i}a s.n., 18008 Granada, Spain
     \and 
     Steward Observatory, University of Arizona, 933 N Cherry Avenue, Tucson, AZ 85721-0065, USA     
   }

   \date{Received ; accepted }
 
  \abstract
   {Dust is formed out of stellar material and is constantly affected by different mechanisms occurring in the ISM. Dust grains behave differently under these mechanisms depending on their sizes, and therefore the dust grain size distribution also evolves as part of the dust evolution itself. Following how the grain size distribution evolves is a difficult computing task that is just recently being overtaking. Smoothed particle hydrodynamic (SPH) simulations of a single galaxy as well as cosmological simulations are producing the first predictions of the evolution of the dust grain size distribution.}  
  {We compare for the first time the evolution of the dust grain size distribution predicted by the SPH simulations with the results provided by the observations. We are able to validate not only the predictions of the evolution of the small to large grain mass ratio ($D_{S}/D_{L}$) within a galaxy, but also to give observational constraints for the recent cosmological simulations that include the grain size distribution in the dust evolution framework.}
 {We select a sample of three spiral galaxies with different masses: M\,101, NGC\,628 and M\,33. We fit the dust spectral energy distribution (SED) across the disc of each object and derive the abundance of the different grain types included in the dust model. We analyse how the radial distribution of the relative abundance of the different grain size populations changes over the whole disc within each galaxy. The $D_{S}/D_{L}$ ratio as a function of the galactocentric distance and metallicity is directly compared to what is predicted by the SPH simulations.}
  {We find good agreement between the observed radial distribution of $D_{S}/D_{L}$ and what is obtained from the SPH simulations of a single galaxy. The comparison agrees with the expected evolutionary stage of each galaxy. We show that the central parts of NGC\,628, at high metallicity and with a high molecular gas fraction, are mainly affected not only by accretion but also by coagulation of dust grains. The centre of M\,33, having lower metallicity and lower molecular gas fraction, presents an increase of the $D_{S}/D_{L}$ ratio, showing that shattering is very effective in creating a large fraction of small grains. Finally, the observational results provided by our galaxies confirm the general relations predicted by the cosmological simulations based on the two grain size approximation. However, we present evidence that the simulations could be overestimating the amount of large grains in high massive galaxies. }
   {}


   \maketitle
%

\section{Introduction}

The physical properties of the dust are directly linked to those of the ISM where it is located. The dust is not only heated by the interstellar radiation field (ISRF) but it is also affected by other mechanisms that are at play in the ISM and can lead to a change in its physical properties and to the destruction of a particular dust grain type. The following processes dominate the evolution of the dust content and the grain size distribution \citep[][]{2015MNRAS.447.2937H}: (i) dust stellar production, (ii) dust destruction by SN shocks in the ISM via sputtering, (iii) grain growth via accretion of metals in the gas phase and (iv) via coagulation\footnote{In case (iii) the dust mass increases, while it stays fixed in case (iv).}; and (v) grain disruption/fragmentation (shattering). All these processes act differently on large and small grains: (i) SNe and AGB stars are predicted to supply mainly large grains\footnote{It is unclear whether the size distribution of grains formed in SN is representative of the grains ejected in the ISM after passage through the reverse shock.} \citep{2007ApJ...666..955N, 2012MNRAS.424.2345V,2017MmSAI..88..397B,2014Natur.511..326G,2019MNRAS.485..440P}, (ii) dust destruction by sputtering affect both large and small grains, with thermal sputtering in SN shocks mostly affecting smaller grains and non-thermal sputtering affecting large grains \citep{2019MNRAS.487.3252H}, (iii) grain growth via accretion is favoured when the number of small grains is large, as small grains have a larger surface-to-volume ratio \citep{2012MNRAS.422.1263H}, (iv) grain growth via coagulation occurs in the dense ISM and moves the grain size distribution towards larger grain sizes \citep{2009A&A...502..845O}, and (v) fragmentation associated to shattering creates a large number of small grains \citep{1996ApJ...469..740J}.

As a result of all these mechanisms, smoothed particle hydrodynamic (SPH) simulations \citep[see e.g.][]{2017MNRAS.466..105A} show that the dust grain size distribution evolves as the galaxy does and the evolution of the total dust mass and the grain size distribution in galaxies are therefore closely related to each other \citep[see also][]{2013MNRAS.432..637A, 2016P&SS..133..107M}. At $t\textless$\,0.1\,Gyr the grain size distribution is dominated by large grains (with radius $a$\,$\gtrsim$\,0.1\,\mi) produced by stars, as the galaxy evolution proceeds shattering occurs and a large number of small grains ($a$\,$\lesssim$\,0.01\,\mi) are formed. Grain growth starts to be important and a bump at $a$\,$\sim$\,0.01\,\mi\ appears between 1-10\,Gyr. The increase of small grains favours the coagulation of small grains into larger ones and consequently the bump is moved towards larger grains at 5-10\,Gyr \citep{2013MNRAS.432..637A,2014MNRAS.440..134A}. The extinction curve strongly depends on the physical and optical properties of dust grains, therefore understanding how the grain size distribution evolves can shed light onto the origin of the differences in the extinction curves among galaxies at high and low redshifts. In particular, since grain growth has the potential to change the grain size distribution, the shape of the extinction curve will change as a result of grain growth \citep{2012MNRAS.422.1263H,2014MNRAS.440..134A}. 
  
At the moment, there are only a few studies incorporating the evolution of the grain size distribution.  \citet{2013MNRAS.432..637A}, suggested that shattering produces a large abundance of very small grains (VSGs) and therefore it enhances the efficiency of grain growth. \citet{2015MNRAS.447.2937H} \citep[see also][]{2019MNRAS.482.2555H} proposed a two-size approximation of large ($a$\,$\gtrsim$\,0.03\,\mi) and small ($a$\,$\lesssim$\,0.03\,\mi) grains and found that the ratio of small to large grains is directly linked to accretion, shattering and coagulation processes. \citet{2017MNRAS.466..105A} used this approximation to produce for the first time hydrodynamical simulations of a galaxy with dust formation and destruction including large and small grains. They were able to trace the evolution of the grain size distribution along the evolution of the galaxy and infer the evolution of the dust extinction curves across time \citep{2017MNRAS.469..870H}.  

Modelling the evolution of each dust grain with a particular size is a very expensive computing task. \citet{2015MNRAS.447.2937H} has proposed a two-size approximation of the grain size distribution that alleviates the computing cost while still describing the main features of the evolution of the grain size distribution (see also \citet{2019arXiv190601917A} and \citet{2019MNRAS.482.2555H} for a recent analysis including a more sophisticated functional form of the grain size distribution). This approach has been very recently applied to cosmological simulations making a breakthrough in the field \citep{2017MNRAS.466..105A,2018MNRAS.478.4905A,2018MNRAS.478.2851M,2019MNRAS.482.2555H,2017MNRAS.469..870H,2019MNRAS.485.1727H}. \citet{2019arXiv190601917A} have been able to follow the full dust grain size distribution for a single galaxy and have confirmed the results provided by the two grain size approximation. Snapshots of the simulations at young ages seems to explain the extinction curves in high-redshift quasars \citep{2017MNRAS.469..870H}. These SPH simulations need to be confronted by observational data, but so far no observational constrains have been available. Comparing the results from the simulations with the observations presented here will allow us to test the assumptions adopted in the simulations. 

We present here a systematic study of the radial variations of the abundance of the different grain types over the galactic disc of three galaxies with different morphological types:  M\,101(SABcd) and NGC\,628(SAcd), two more massive spiral galaxies; and M\,33(SAc), which is a less massive spiral galaxy. In the future we plan to extend this study to a larger sample of galaxies covering a wider morphological range. We have derived ratios of small to large grains at different galactocentric distances covering a wide range in metallicity (12+log(O/H)\,$\sim$\,7.8-8.8).  The variation of the relative abundance of the different grain types within the galactic discs sets up restrictions on the results of the SPH simulations which incorporate dust formation and destruction within a two grain size formalism. We make a step forward in applying the same fitting methodology to a wide sample of galaxies from the KINGFISH \citep{2011PASP..123.1347K} and Dwarf Galaxy Survey \citep[DGS,][]{2013PASP..125..600M} and compare the observational results with the recent cosmological simulations. This comparison provide an observational benchmark for future simulations. 

In section\,\ref{sec:data} we present the data and fitting technique applied to the observations. The results of the fitting across the disc of our three galaxy sample is compared with the SPH simulations of a single galaxy in section\,\ref{sec:comp}. In section\,\ref{sec:S2L_prop}, we compare the predictions of cosmological simulations with the results from fitting the integrated spectral energy distributions (SEDs) of a large sample of galaxies. We discuss the main agreements and disagreements between observations and simulations in section\,\ref{sec:disc}. The conclusions are presented in section\,\ref{sec:con}.
 
\section{Data and SED fitting}\label{sec:data}

In this study, we analyse the data in two different ways: one based on spatially resolved analysis for three individual galaxies,  and the other which analyses the integrated SED of a sample of galaxies. The first approach will allow us to study how the relative contributions of the different grain types changes with metallicity and radius {\it within} the galactic disc; the second approach gives information about the relative grain size distributions of galaxies as a whole, and therefore is useful when comparing the results of the evolution of the grain size distribution in a cosmological volume. In this last approach we have included galaxies from two different surveys in order to cover a wide range of morphological types and other physical properties of galaxies, such as metallicity, mass and star formation rate (SFR). 

\subsection {Spatially resolved studies}

We use infrared data from the KINGFISH collaboration \citep{2011PASP..123.1347K} for M\,101 and NGC\,628, and from HerM33es \citep{Kramer:2010p688} for M\,33, to derive dust mass maps over the discs of these galaxies. We fit the infrared spectral energy distributions from 3.6\,\mi\ to 500\,\mi\ at each location of the disc of these spiral galaxies on a pixel-by-pixel basis. We take only those pixels with reliable SEDs, i.e. pixels with fluxes in all bands from 3.6\,\mi\ to 500\,\mi\  with values above 3 times the standard deviation of the background value in each filter. We apply another mask to cover a wide area of the disc with diffuse and star forming regions while avoiding low surface brightness areas that would lead to unreliable fits. 

\begin{figure} 
\centering
 \includegraphics[width=0.5\textwidth]{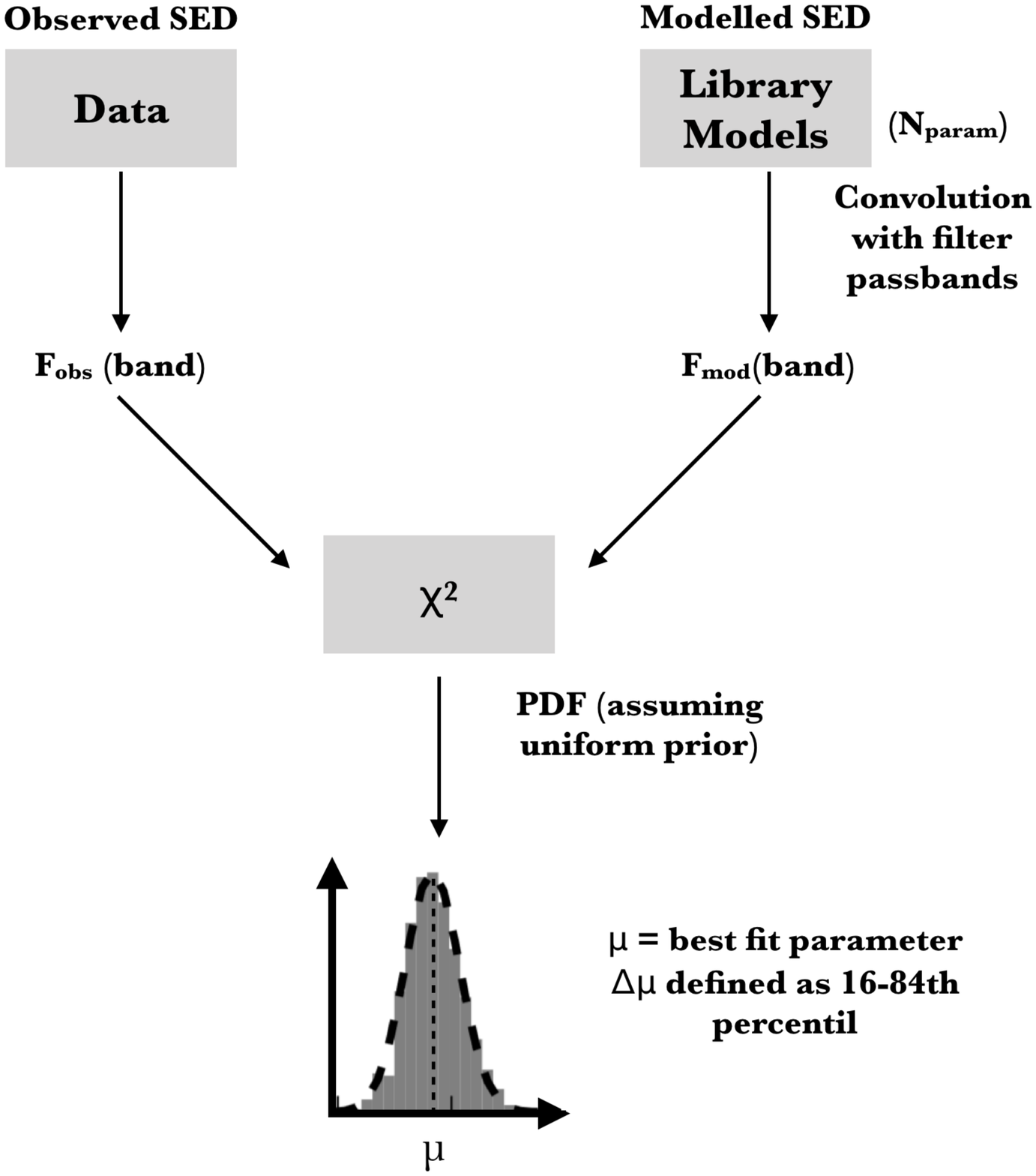}
 \vspace{-1cm}
   \caption{Schematic representation of the SED fitting procedure performed in this study. N$_{\rm param}$ is the number of free parameters assumed in the models: relative mass fraction of PAHs, VSGs and BGs, the scale factor of the ISRF, and a scale factor to describe the contribution of the old stellar population to the NIR-continuum. PDF is the probability density function obtained assuming an uniform prior distribution of the initial parameters. F$_{\rm mod}$(band), and F$_{\rm obs}$(band) are the corresponding modelled and observed fluxes in each band.}
   \label{fig:SEDfit_sketch}
\end{figure}

We use the classical \citet{1990A&A...237..215D} dust model, which consists of three different grain populations: polycyclic aromatic hydrocarbons (PAHs), VSGs and big silicate grains (BGs), and assume an ISRF with the shape as the solar neighbourhood given in \citet{Mathis:1983p593}. The fitting procedure uses a Bayesian approach (see Fig.\,\ref{fig:SEDfit_sketch} for a schematic representation of the procedure): we first create a library of models with different values of the input parameters, covering a wide range of possible solutions for each galaxy. Convolution of the modelled SEDs with the corresponding filter bandpass of our observations allows us to obtain the fluxes in each band for each model in the library. With the observed and modelled fluxes we compute the $\chi^2$ value associated to each model. We then build up the probability density function to obtain the best parameter value and its uncertainty. The fitting procedure gives us not only the dust mass but also the relative mass fraction for each grain type, as well as the scale factor of the ISRF. With this information we create maps of the dust abundance for each grain type across the discs of our three galaxy sample. Using HI and CO observations we also produce gas-to-dust mass ratio maps of the disc of each individual galaxy. A detailed explanation on the fitting procedure and the derivation of the gas-to-dust ratio mass maps is presented in \citet{2019MNRAS.483.4968V} for M\,101 and NGC\,628 and \citet{2018A&A...613A..43R} for M\,33. 

\subsection {Integrated SEDs of a galaxy sample}
We apply the same fitting procedure to the integrated SEDs of the KINGFISH  \citep{2011PASP..123.1347K} and DGS \citep{2013PASP..125..600M} surveys. The integrated photometry of the KINGFISH and DGS is presented in \citet{2017ApJ...837...90D} and \citet{2015A&A...582A.121R}, respectively. To avoid contamination of spurious fits we only take galaxies with detections in all bands, eliminating uncompleted SEDs and SEDs with upper limits. The strategy for the fitting procedure is the same as the one performed for the disc of M\,101, NGC\,628 and M\,33, but we change the range for the ISRF scale to provide a better coverage of the possible solutions in the case of integrated galaxies. We furthermore remove those galaxies having fits with residuals in the 24\,\mi\ and 70\,\mi\ bands above 35\%.    

The \citet{1990A&A...237..215D} dust model assumes that PAHs are grains with radii  (0.4-1.2)$\times$10$^{-3}$\mi, VSGs correspond to grains with radii (1.2-15)$\times$10$^{-3}$\mi, and BGs those grains with radii larger than 15$\times$10$^{-3}$\mi. The separation in sizes implies that, to compare with the simulations, small grains are PAHs and VSGs and large grains corresponds to silicates BGs in our observations. Despite of the existence of new dust models more sophisticated than the classical one proposed in \citet{1990A&A...237..215D} we have decided to apply this last model for the present study. The main reason is its simplicity and the small number of free parameters compared to more sophisticated recent dust models \citep[e.g.][]{2007ApJ...657..810D,2011A&A...525A.103C,2013A&A...558A..62J}. In \citet{2016A&A...595A..43R} we compared both \citet{1990A&A...237..215D} and \citet{2011A&A...525A.103C} dust models and found that they both agree in reproducing the relative abundance of VSG and BG grains.

\section{Comparison with SPH simulations of a single galaxy}\label{sec:comp}
\citet{2017MNRAS.466..105A} and \citet{2017MNRAS.469..870H} performed GADGET-3 SPH simulations (originally explained by \citet{2005MNRAS.364.1105S}) of an isolated galaxy incorporating dust evolution mechanisms. The simulations are based on those provided by the low-resolution model of AGORA simulations \citep{2014ApJS..210...14K}, with a spatial scale of $\sim$\,80\,pc. An initial disc is created as a characteristic one of galaxies at z\,$\sim$\,1. The old stellar populations are dynamically collisionless star particles that build the initial mass of the disc and bulge. The initial physical parameters of the simulated galaxy are shown in Table\,1 in \citet{2017MNRAS.466..105A} with initial stellar and gas distributions described in \citet{2014ApJS..210...14K}. Once the galaxy is built with the initial disc, star formation occurs following the prescription given in Eq.\,2 of \citet{2017MNRAS.466..105A}, which parameterises the SFR in terms of the gas density. Each new created star particle is seen as a single stellar population with Chabrier (2003) initial mass function (IMF) and carries information of the stellar mass, metallicity and formation time. 

Dust is produced by the metals ejected by SNe and the metal enrichment is assumed to occur $\sim$\,4\,Myr after the star formation. Dust production by AGB stars is not included in these simulations. The amount of dust produced by AGB stars is still in debate, as well as how the dust yields from AGB stars depend on other physical parameters such as metallicity, star formation history or internal parameters of the stars \citep[e.g.][]{2018MNRAS.473.5492N,2019MNRAS.487..502N}. Contribution to the dust budget from AGB stars can exceed the amount of dust produced by SNe at later ages, with a higher contribution of AGB stars at lower metallicity \citep[e.g. Fig. 3 in ][]{2009MNRAS.397.1661V}. However, accurate dust budget calculations done in the Magellanic Clouds \citep{2016MNRAS.457.2814S,2013A&A...555A..99Z}, where stars can be individually observed,  suggest that the amount of dust observed in the ISM is higher than the cumulative dust mass produced by AGB stars during cosmic time scales. We therefore consider that not including AGB star dust production in the simulation is not affecting severely the results of the simulations. Future observational facilites are needed to estimate with accuracy the amount of dust produced by AGB stars in galaxies located a further distance than the Magellanic Clouds \citep[e.g.][]{2015ApJS..216...10B}.

The simulations adopt the two grain size distribution approximation presented in  
\citet{2015MNRAS.447.2937H}, separating the grain size distribution into small and large grains at $a\sim0.03\,\mi$ (with $a$ being the grain radius). The separation is justified based on the full calculation of the grain size distribution performed by \citet{2013MNRAS.432..637A}, who showed that the processes dominating the small grain abundance create a bump in the size distribution at small grain sizes while dust production by stars creates a bump in the large grain size regime. The separation between these two bumps is at $a\simeq0.03\,\mi$. The \citet{1990A&A...237..215D} dust model assumes a separation between small and large grains of $a=0.015\,\mi$, which is similar to the separation limit of the two grain size approximation and also separates the two bumps predicted by the analysis of the full grain size distribution. The time evolution of the dust abundances is followed in each gas particle and includes stellar dust production, SN destruction, grain disruption by shattering in the diffuse ISM, and grain growth by coagulation and accretion in the dense ISM. The simulations assume a solar metallicity of Z$_{\odot}$\,=\,0.02 and therefore, for consistency, we will assume the same value for the observations presented here. We refer the reader to \citet{2017MNRAS.466..105A} for a full description on how each process is incorporated into the simulations and the dependencies of the time scales, and to \citet{2017MNRAS.469..870H} for a study of the extinction curves predicted by these simulations. 

\subsection{Radial profiles of the dust mass}\label{sec:Mdust_rad}

\begin{figure} 
\includegraphics[width=0.5\textwidth]{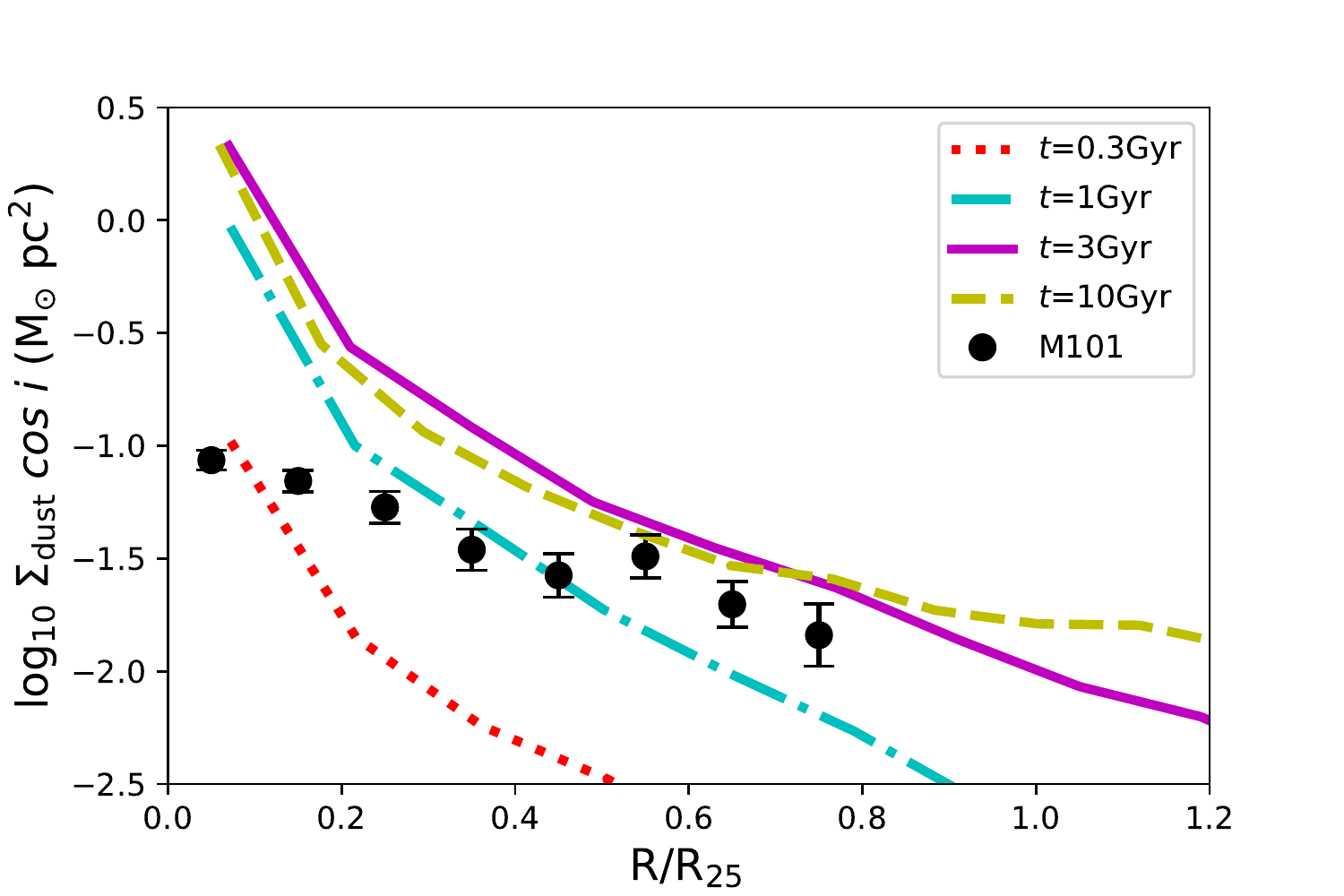}
  \includegraphics[width=0.5\textwidth]{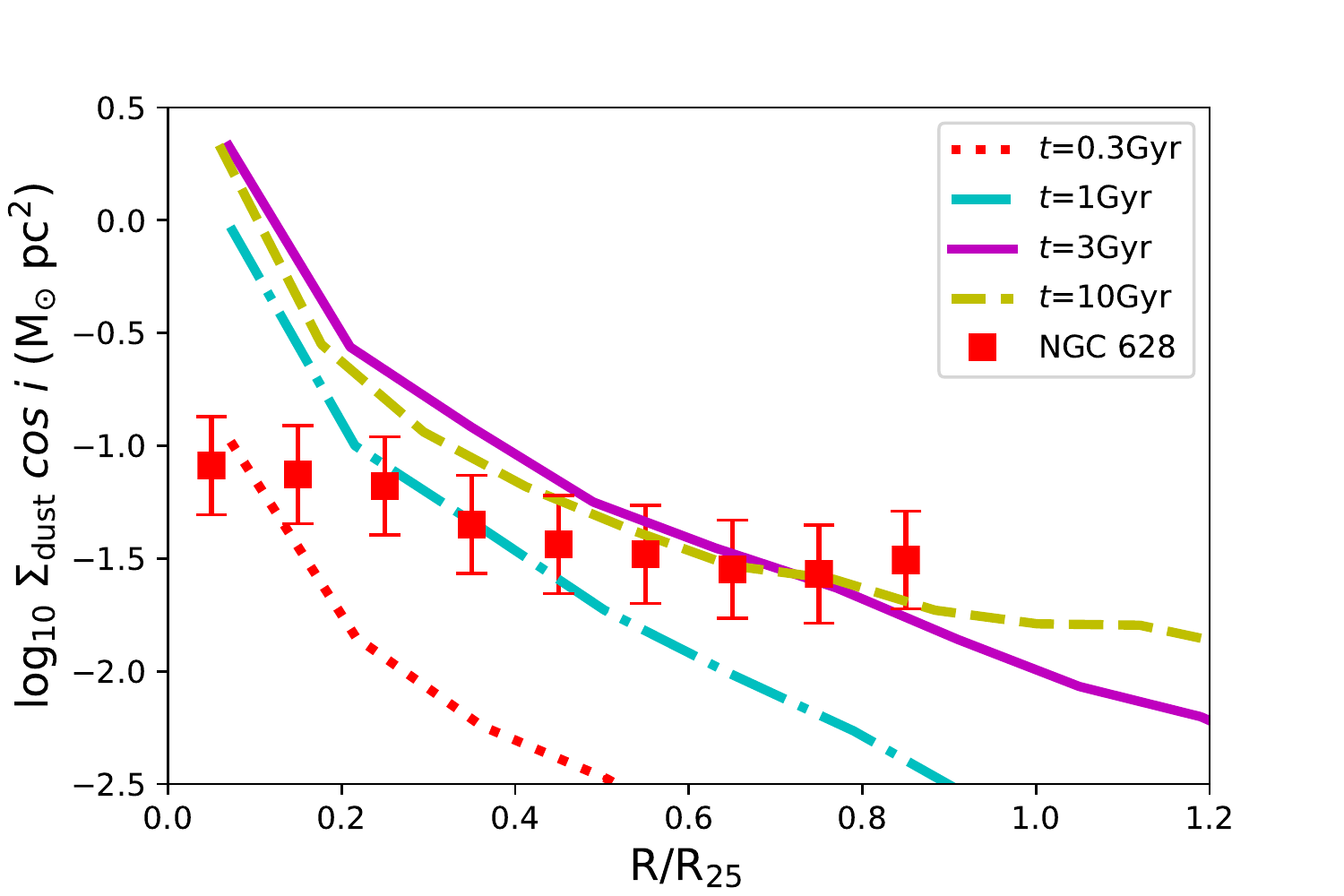}
 \includegraphics[width=0.5\textwidth]{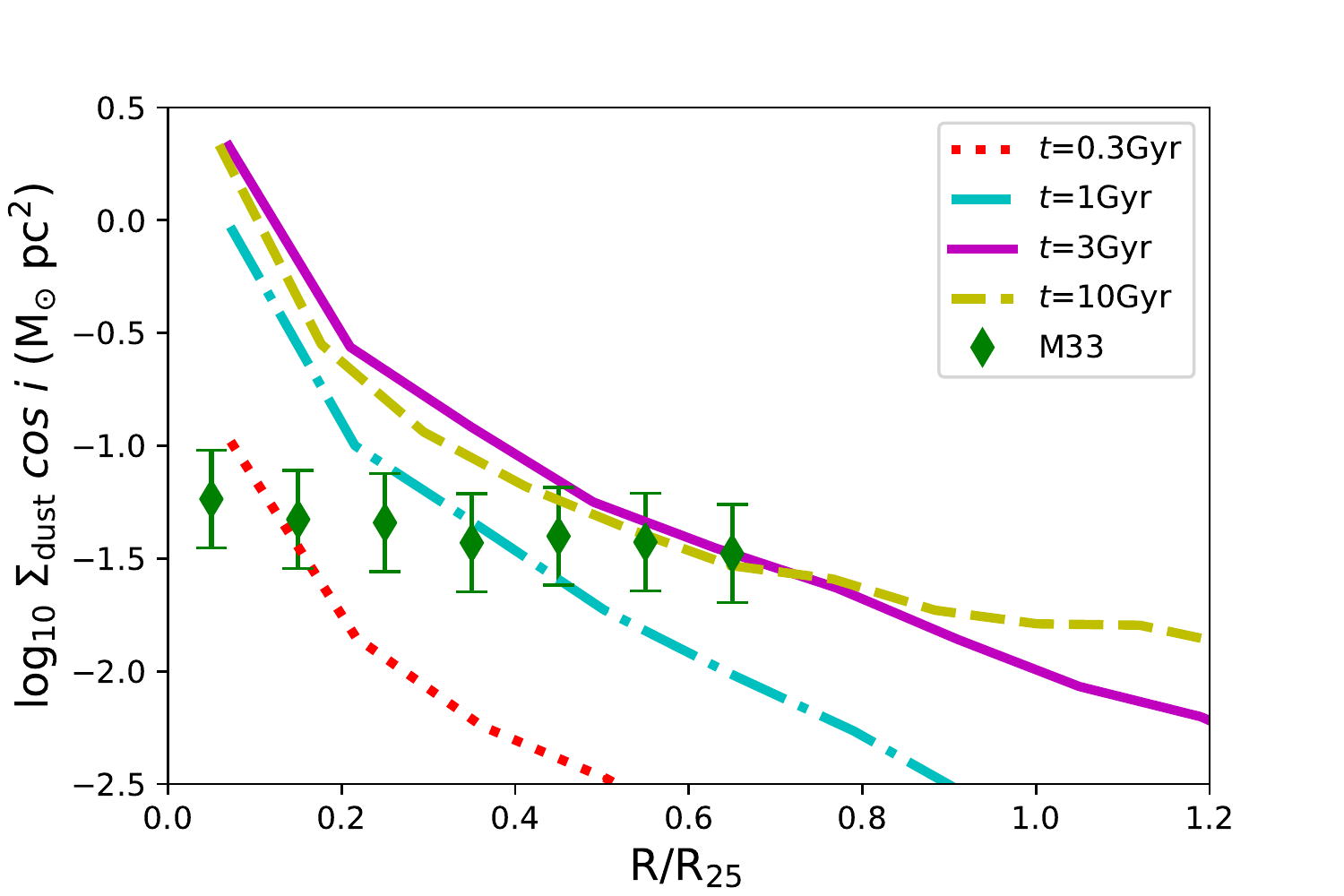}
   \caption{Radial profile of the dust mass surface density for M\,101 (top), NGC\,628 (middle), and M\,33 (bottom) compared to the profiles provided by the simulations performed by \citet{2017MNRAS.466..105A} and \citet{2017MNRAS.469..870H}. Dust production in stellar ejecta, dust destruction by SN shocks, grain growth by accretion and coagulation, as well as shattering, are taking into account in the simulations.}
   \label{fig:SDMdust_Hou17}
\end{figure}

We first compare the radial variation of the dust content predicted by the simulations with the results from the observations. Fig.\,\ref{fig:SDMdust_Hou17} shows the radial profiles of the dust mass surface density obtained from the observations as well as the predictions of the simulations at different time steps of the galaxy evolution. The data points  shows the mean value within elliptical rings of 0.1\,R$_{\rm 25}$ width and the error bars represent the variation of the dust mass surface density in each ring. In general the dust mass surface density decreases towards the outer parts of the galaxy. The decrement is more pronounced for M\,101 and NGC\,628 than for M\,33, as the first two galaxies show higher metallicity values in their centres and more pronounced metallicity gradients: 12+log(O/H)\,=\,8.72\,--\,0.83\,R/R$_{\rm 25}$ for M\,101; \citep{2016ApJ...830....4C}, 12+log(O/H)\,=\,8.84\,--\,0.49\,R/R$_{\rm 25}$ for NGC\,628 \citep{2015ApJ...806...16B}; and 12+log(O/H)\,=\,8.50\,--\,0.38\,R/R$_{\rm 25}$ for M\,33 \citep{2011ApJ...730..129B} . 

The model at 0.3\,Gyr shows very low dust masses compared to the observations, while models at later ages, 1-10\,Gyr, fit at least part of the disc of the galaxies. The simulations predict however high dust masses and steeper radial gradients in the centre of the galaxy at all ages, which do not agree with the mild gradients across the whole disc obtained from the observations. The discrepancy might be produced by the assumptions in the initial conditions of the simulations. Dust is produced in the simulations by the metals ejected from the core-collapse SNe. The SN rate depends on the IMF and the SFR prescription assumed in the simulations. The SFR is parameterised in terms of the gas density \citep[see Eq.\,2 in][]{2017MNRAS.466..105A} and therefore the metal enrichment is influenced by the initial gas distribution of the galactic disc. The fact that in the centre of the galaxies the simulations do not describe properly the observations would indicate that the gas distribution of the simulated galaxy does not completely reproduce the exact gas distribution of our three galaxies. Indeed, the simulated galaxy has a large bulge, while none of these galaxies shows signatures of large bulges in their centres: \citet{2010PASP..122.1397S} classified NGC\,628 and M\,101 as galaxies hosting a pseudobulge, and \citet{2007ApJ...669..315C} did not find any kinematical signature of a bulge in M\,33. We should point out here that our goal in the present paper is not to perform a SPH simulation of each individual galaxy but to provide a global observational comparison that tests the results and provide clues for possible caveats of the simulations. Some galaxies in the KINGFISH survey present signatures of central bulges \citep[e.g. NGC~4536 and NGC~1291,][]{2015ApJS..219....4S}. In the future, we plan to extend our spatially resolved study to a larger number of galaxies from the KINGFISH survey, covering a broader range of morphologies. In this way, we will be able to test whether our interpretation of the disagreement of the radial profile between data and simulations is correct and verify whether galaxies with a larger bulge indeed fit the simulations better.

\subsection{Dust to gas mass ratio versus metallicity and radius}\label{sec:comp_D2G}

\begin{figure*} 
 \includegraphics[width=.5\textwidth]{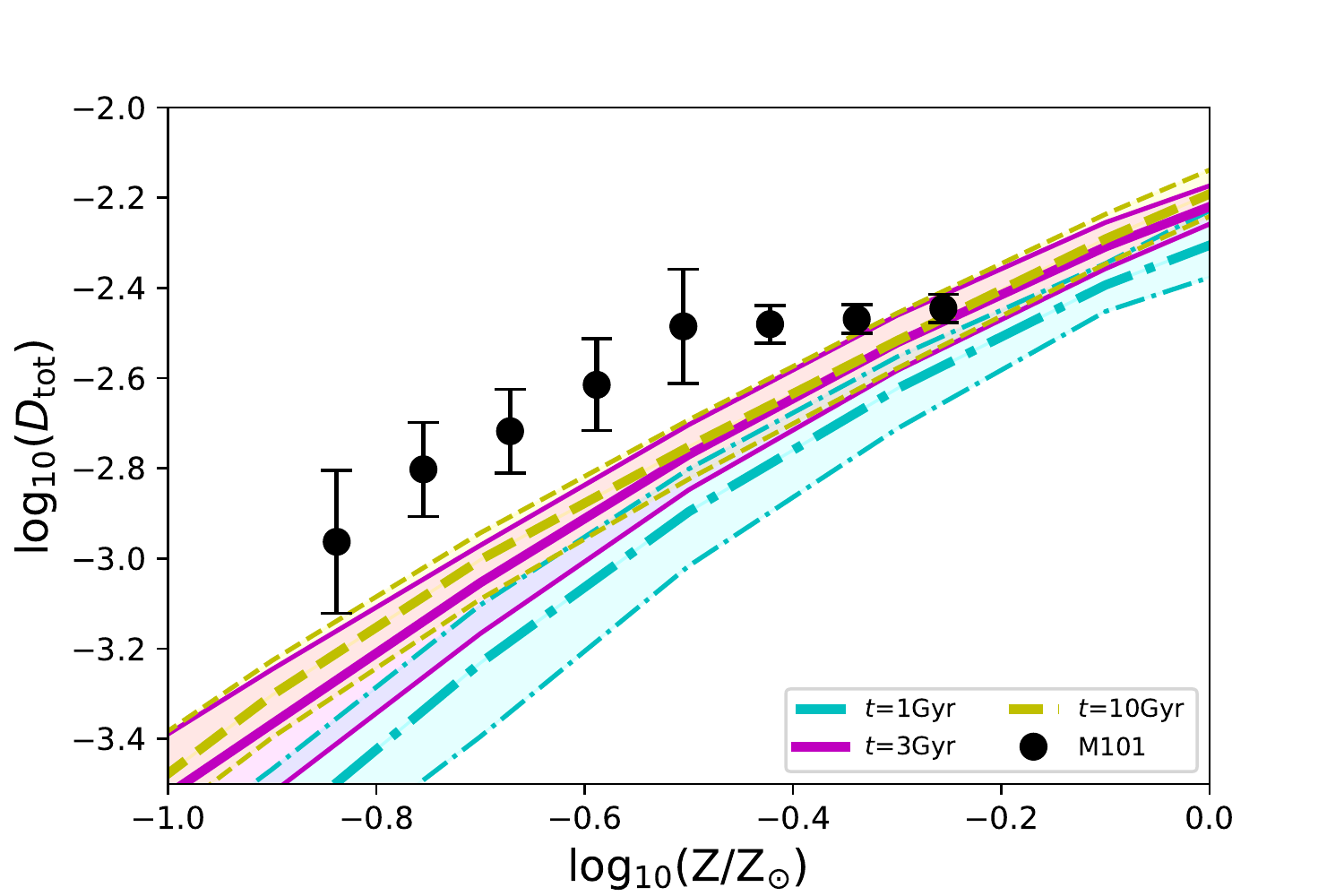}
    \includegraphics[width=.5\textwidth]{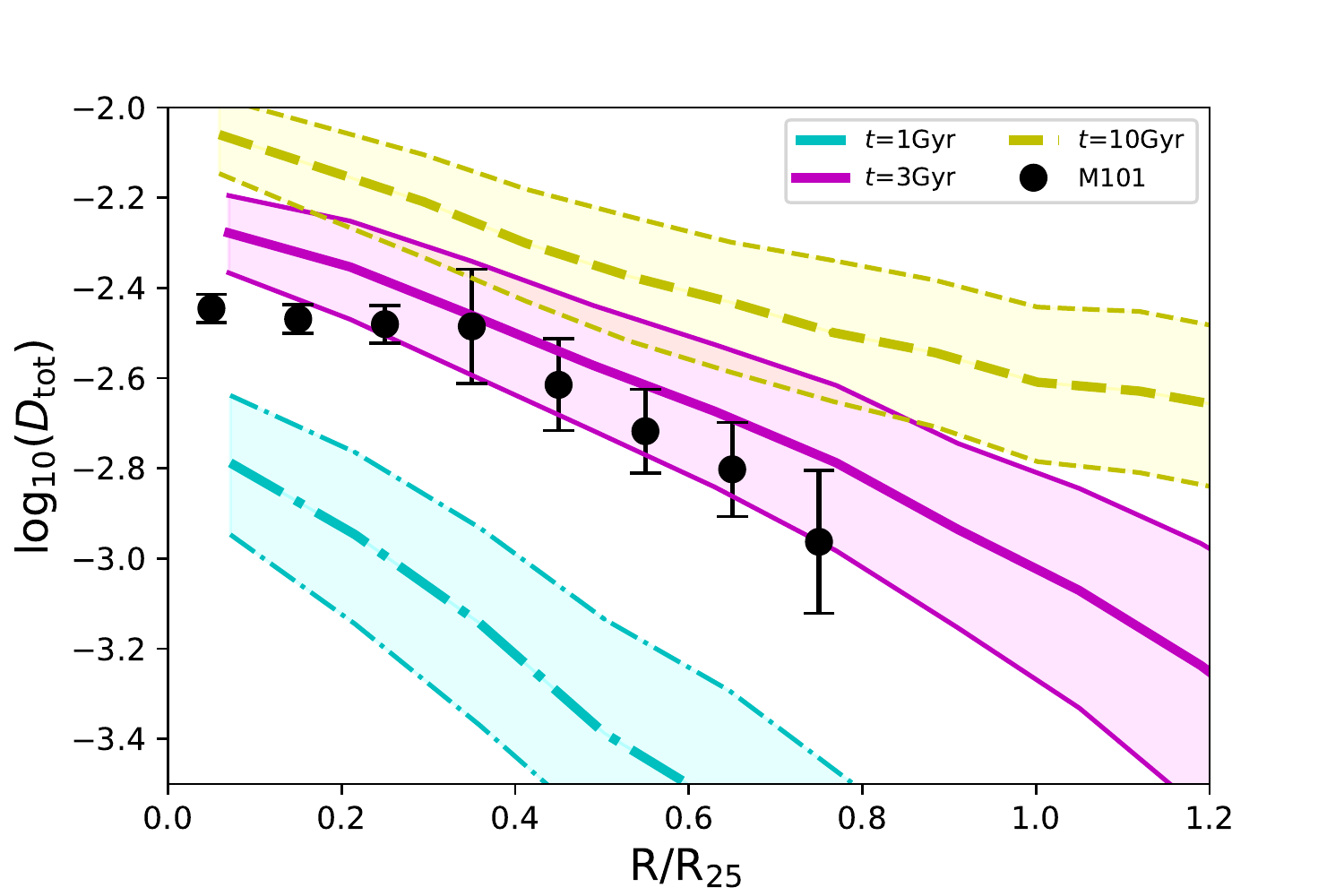} 
  \includegraphics[width=.5\textwidth]{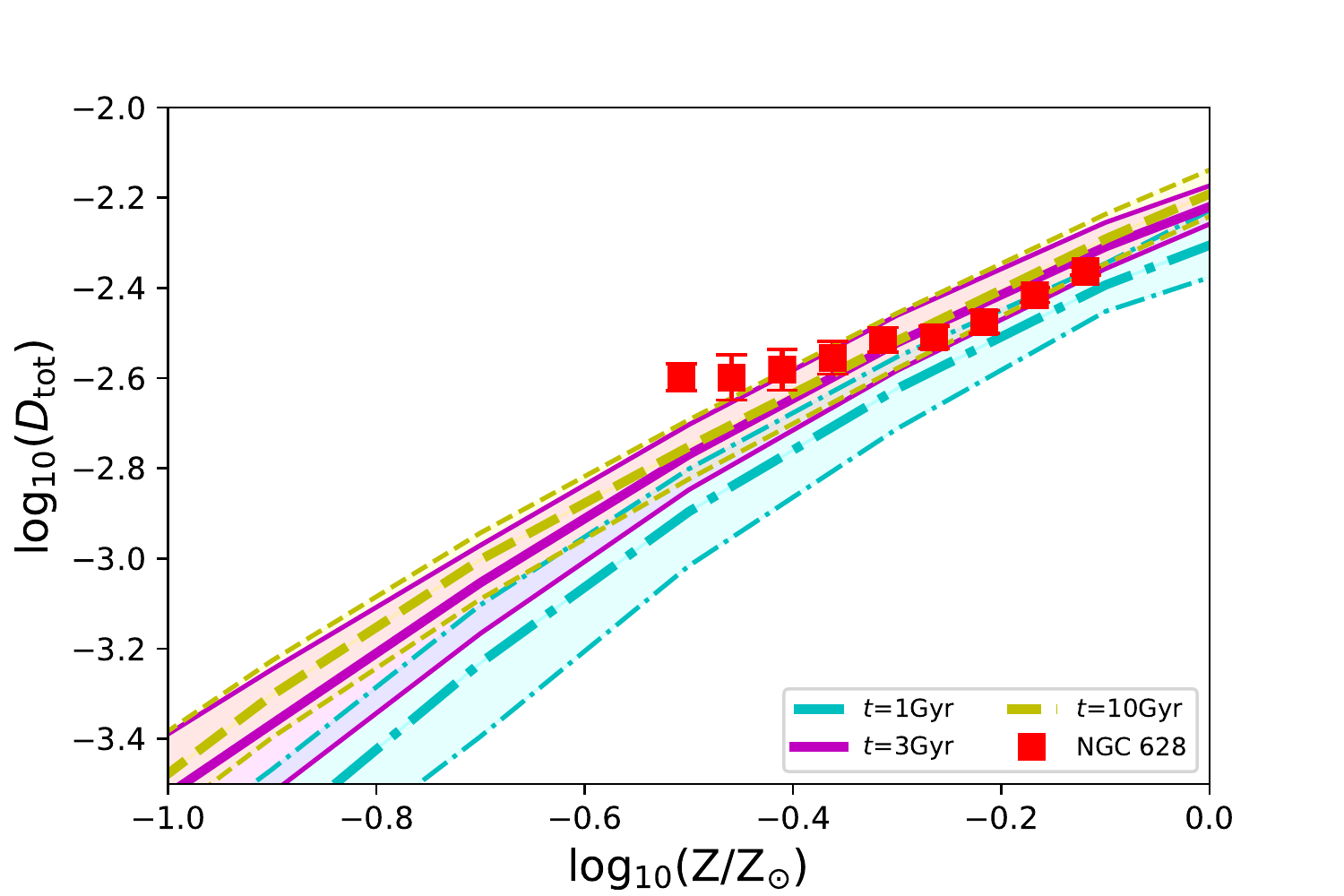}
        \includegraphics[width=.5\textwidth]{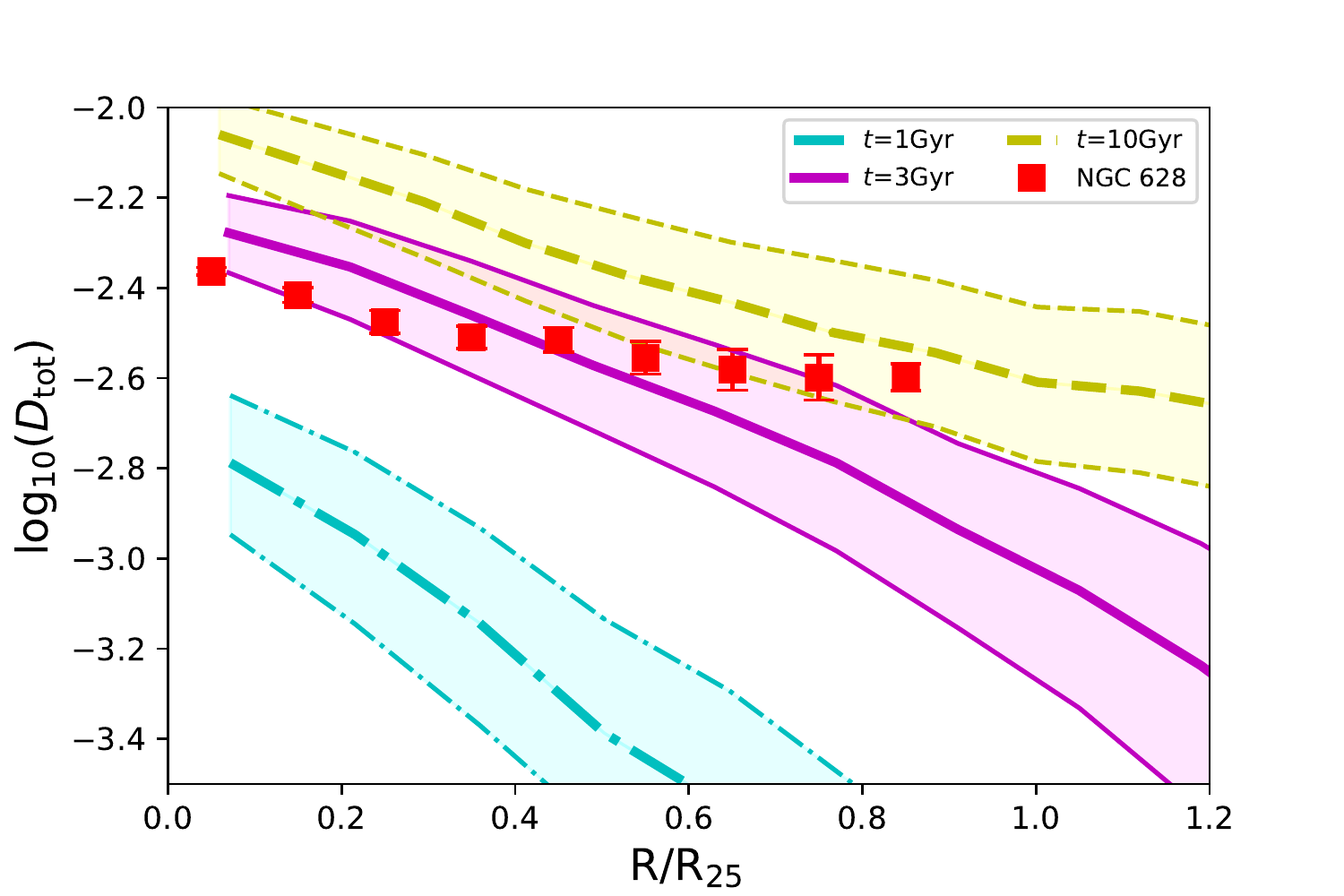}
 \includegraphics[width=.5\textwidth]{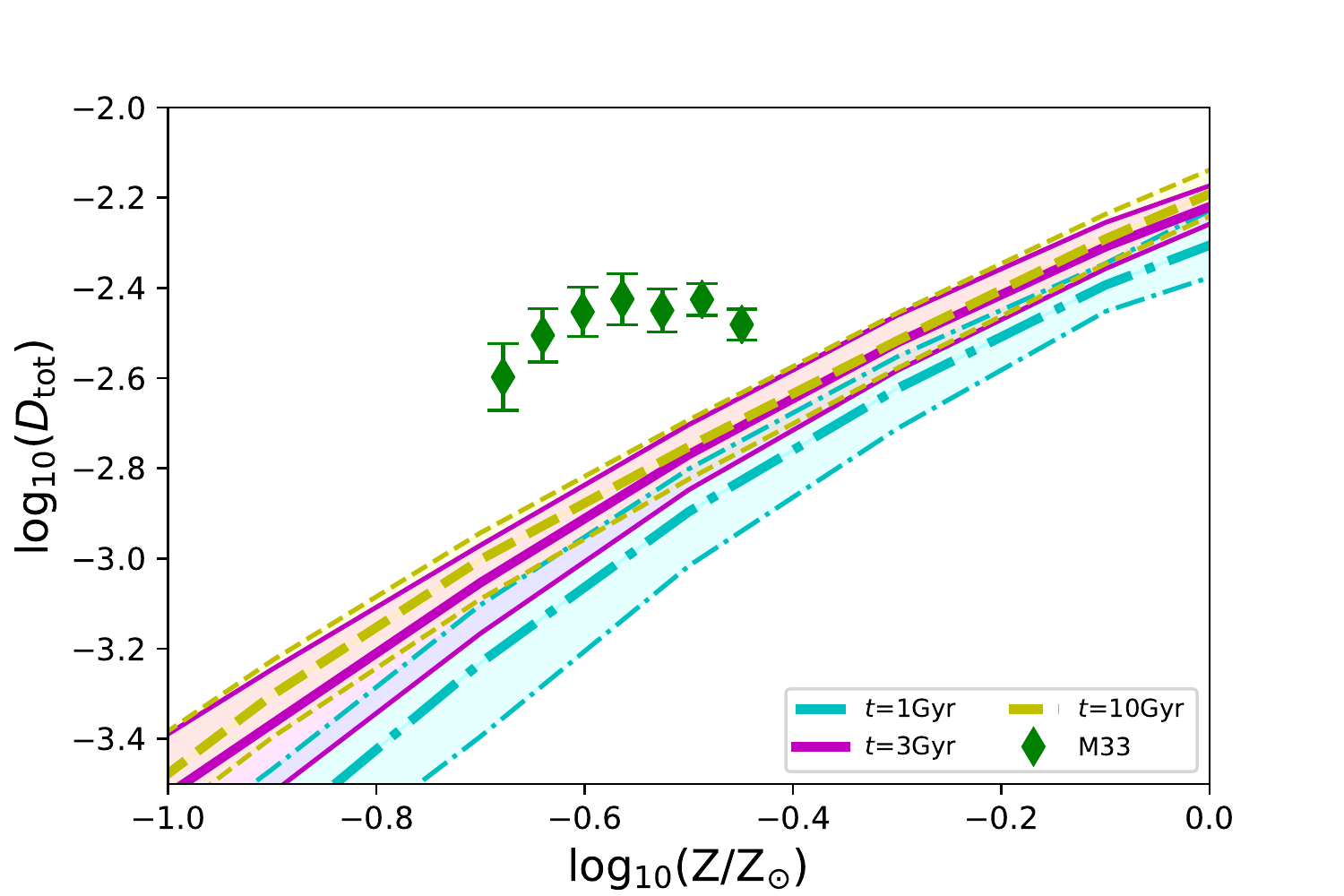}
   \includegraphics[width=.5\textwidth]{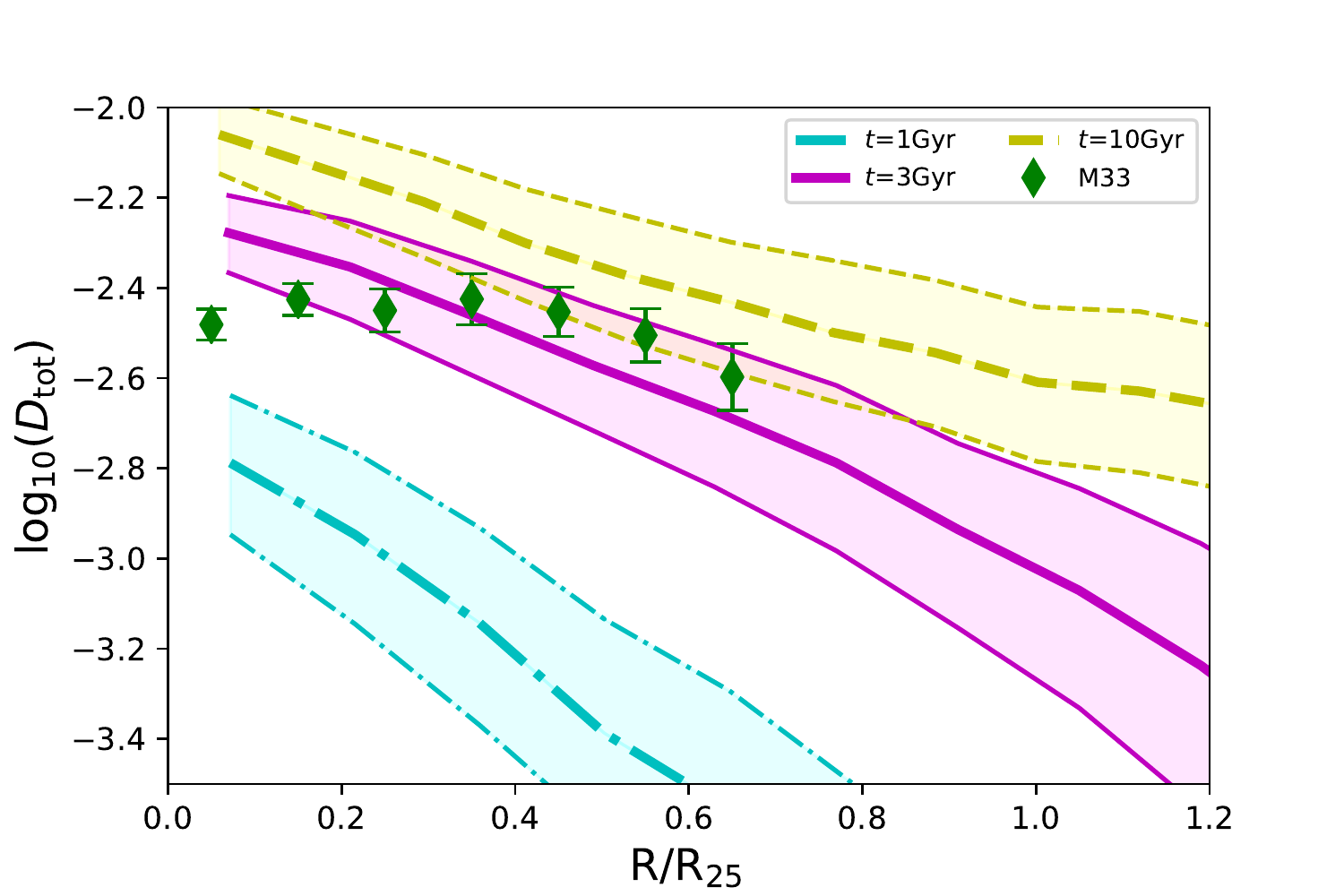}
   \caption{Comparison of the variation of the dust-to-gas mass ratio ($D_{tot}$) with metallicity (left) and galactocentric radius (right) for M\,101 (top), NGC\,628 (middle), and M\,33 (bottom) with the SPH simulations of an isolated galaxy performed by \citet{2017MNRAS.466..105A} and \citet{2017MNRAS.469..870H}. Thick lines represent mean values for each model in the simulation and the coloured areas cover the 25 and 75 percentile of the particle distribution. A solar metallicity of Z$_{\odot}$\,=\,0.02 has been adopted for our observations for consistency with the solar metallicity assumed in the simulations.}
   \label{fig:D2Gcomp_Hou17}
\end{figure*}

In Fig.\,\ref{fig:D2Gcomp_Hou17} we show how the dust-to-gas mass ratio ($D_{tot}$) derived from our observations compares to the SPH simulations. In the right panels we show the radial variation of $D_{tot}$ (normalised to R$_{25}$), and in the left panels we show the variation of $D_{tot}$ as a function of metallicity. The observed data points have been derived using the gas-to-dust mass maps obtained previously in \citet{2019MNRAS.483.4968V} for M\,101 and NGC\,628, and \citet{2018A&A...613A..43R} for M\,33. We integrate these maps in elliptical rings of 0.1\,R$_{\rm 25}$ width and take the mean value as the representative value for the gas-to-dust mass ratio in each ring. Each data point in Fig.\,\ref{fig:D2Gcomp_Hou17} is therefore the inverse of the mean value of each elliptical ring and the error bar represents the corresponding variation within the ring. The metallicity in each ring (scaled to Z$_{\odot}$\,=\,0.02) has been obtained using the metallicity gradients derived from spectroscopic observations for each galaxy: \citet{2016ApJ...830....4C} for M\,101, \citet{2015ApJ...806...16B} for NGC\,628, and \citet[][]{2011ApJ...730..129B} for M\,33 (see previous section). 

Models at 0.3\,Gyr are not presented in these figures as they are not able to reproduce our observations. This seems plausible as these galaxies are already older than 0.3\,Gyr: \citet{2013ApJ...769..127L} and \citet{2014MNRAS.437.1534S} estimated ages for the stellar population of the disc of $\tau$\,$\sim$6-7\,Gyr for M\,101 and $\tau$\,$\sim$3-6\,Gyr for NGC\,628; while for M\,33 \citet{2017MNRAS.464.2103J} shows that the oldest star formation epoch occurred at $\tau$\,$\approx$3-6\,Gyr with a second epoch $\approx$200-300\,Myr ago, reaching a level of almost four times the SFR in the earlier epoch.

The agreement between $D_{tot}$ derived from observations and the predictions of the simulations is good at high metallicities and in the central parts of the galaxies, but at low metallicities the observed $D_{tot}$ is always higher than the values predicted by the simulations. Since the dust mass surface density is not well reproduced by the observations, the fact that $D_{tot}$ follows the behaviour of the observations in the central parts of the galaxies indicates that the gas mass fraction assumed in the simulations might be too high. An excess in the gas mass has already been suggested in \citet{2019MNRAS.485.1727H}, as their simulations seem to produce a few times higher gas-to-stellar mass ratio than the observed ratio derived in \citet{2017ApJS..233...22S}. There is also an excess of the simulated gas mass compared to the results of the ALFALFA survey presented in \citet{2015MNRAS.447.1610M}.

In the lower metallicity range, corresponding to the outer parts of M\,101 and M\,33, the observed $D_{tot}$ is not well reproduced by the SPH simulations. Only the behaviour of $D_{tot}$ in NGC\,628, which has a relatively high molecular gas mass fraction across its disc and does not reach as low metallicities as the other two galaxies, seems to agree with the simulations. The simulations seem to fail to explain the $D_{tot}$ behaviour at low metallicities and low dense gas fractions.

The comparison between simulations and observations is better in the $D_{tot}$ versus radius relation (Fig.\,\ref{fig:D2Gcomp_Hou17}, right panels)  than in the relation with metallicity (Fig.\,\ref{fig:D2Gcomp_Hou17}, left panels)  . This is somehow expected: binning in metallicity follows exactly the metal enrichment of each particle in the simulations, which is subject to the initial conditions of the simulation: SFR, IMF and initial gas distribution in the disc, as well as the metal yields assumed in the simulations \citep[see][for details]{2014ApJS..210...14K}. In our observations we are assuming a metallicity radial gradient derived from optical spectroscopic data where the electron temperature has been previously obtained (see Section\,\ref{sec:Mdust_rad}). In this sense we are not taking into account possible azimuthal metallicity variations within the disc, each radial bin has a single metallicity defined by the metallicity gradient. However, it is expected that these azimuthal variations, when present, are not above $\sim$\,0.1\,dex typically. The fact that the simulations are able to reproduce well the radial trends of the observations and not the metallicity might indicate that azimuthal variations in metallicity predicted by the simulations are large and significantly affected by the initial conditions.

\subsection{Small to large grain ratio versus metallicity and radius}\label{sec:S2L_rad}

\begin{figure*} 
 \includegraphics[width=.5\textwidth]{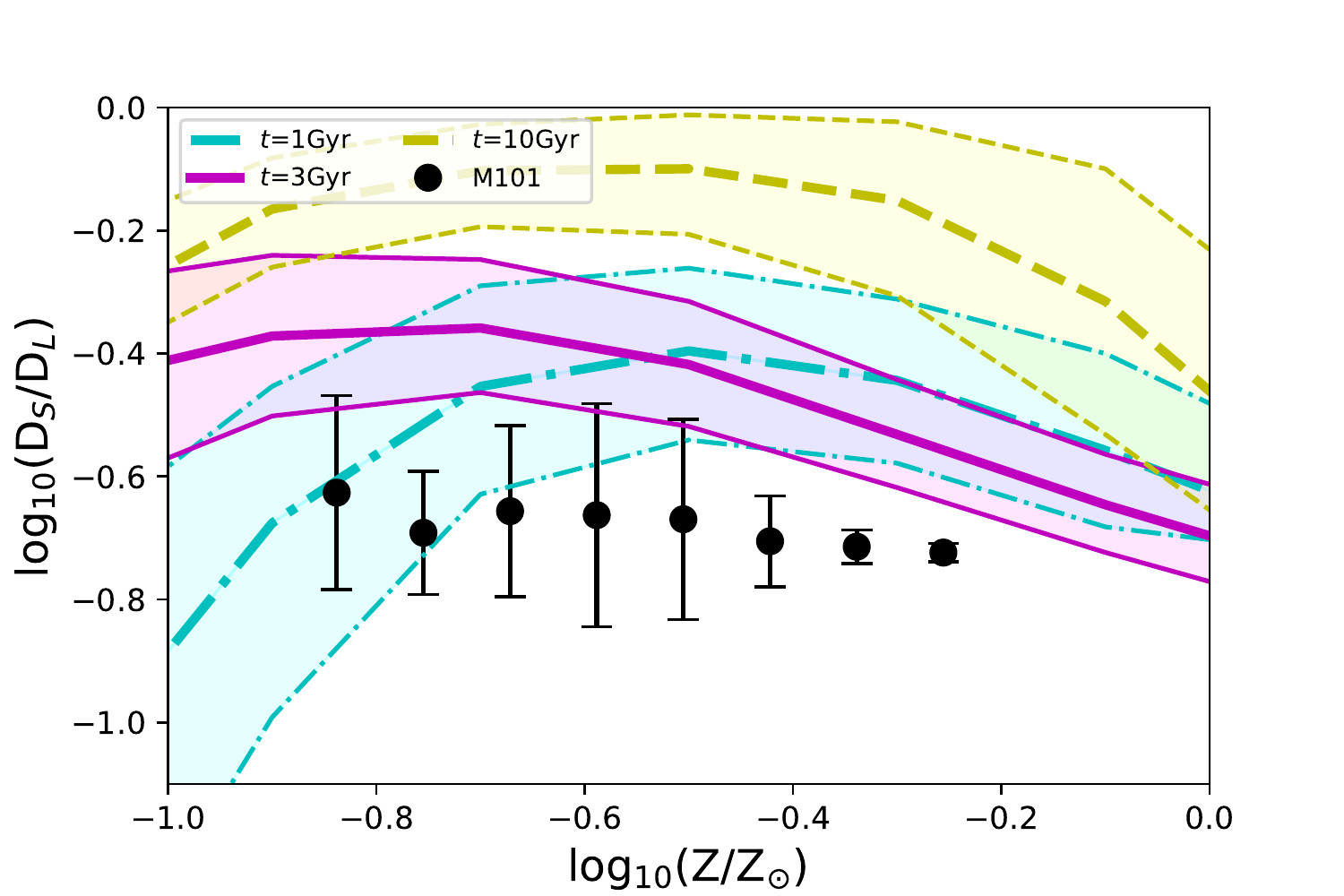}
    \includegraphics[width=.5\textwidth]{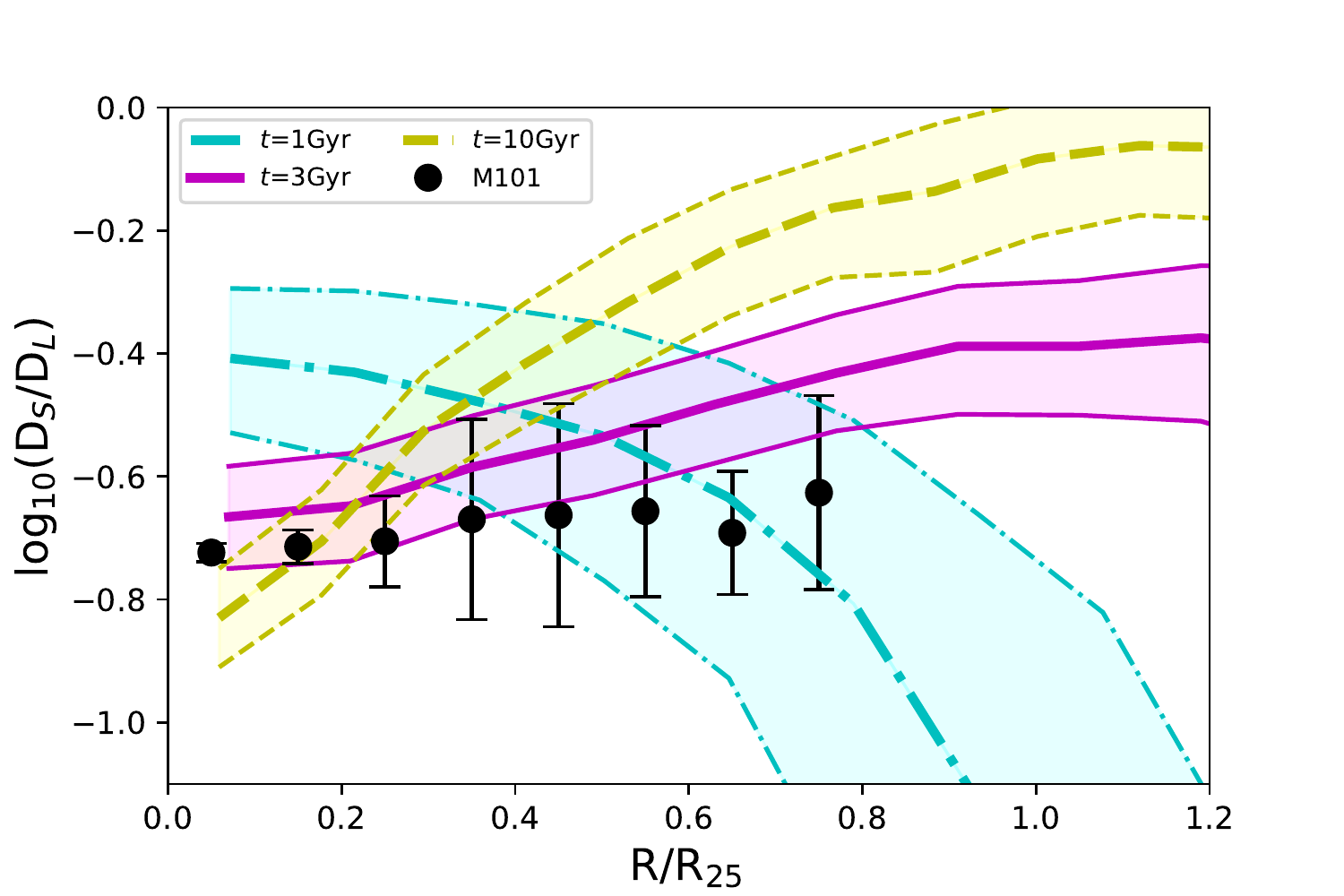}

  \includegraphics[width=.5\textwidth]{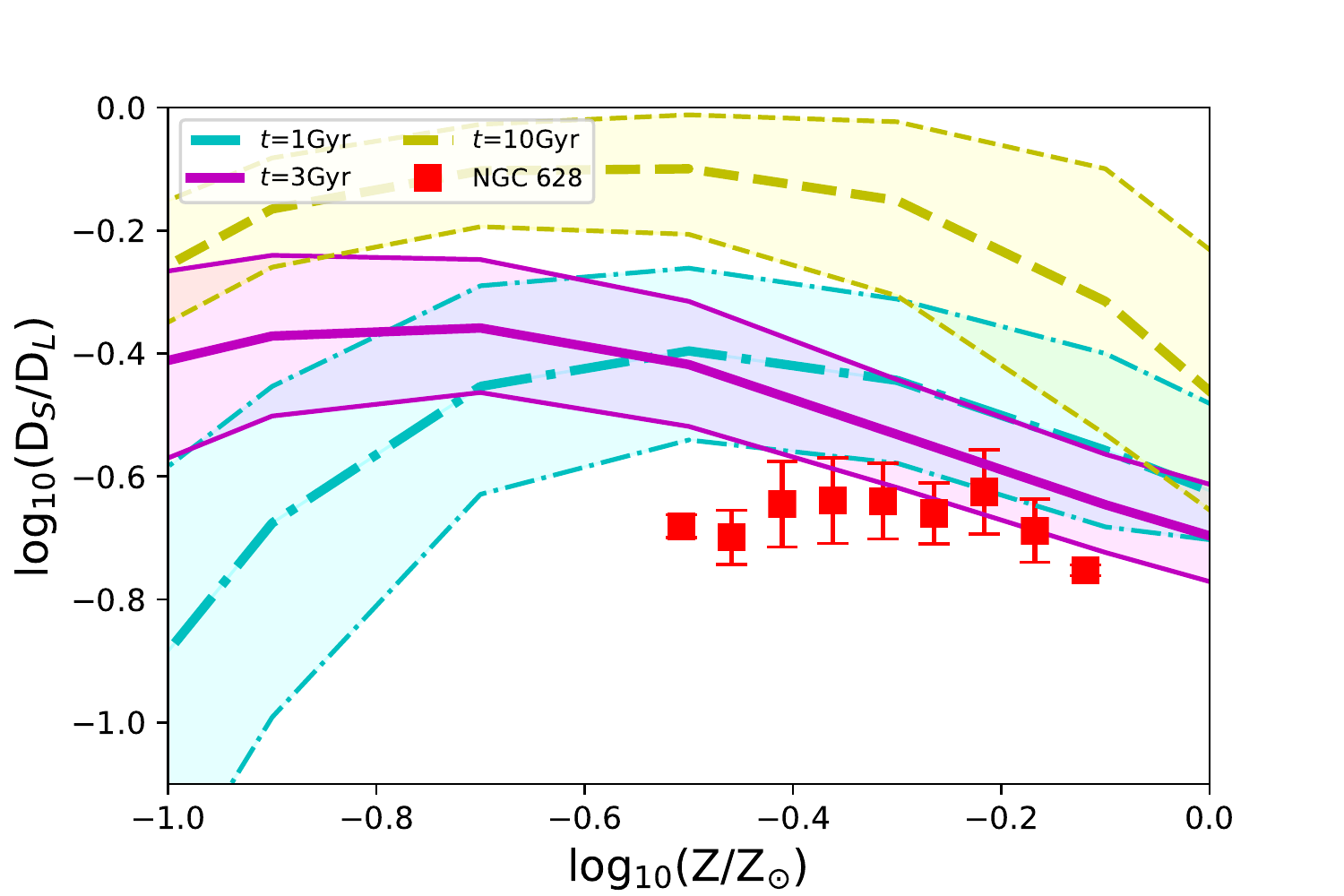}
  \includegraphics[width=.5\textwidth]{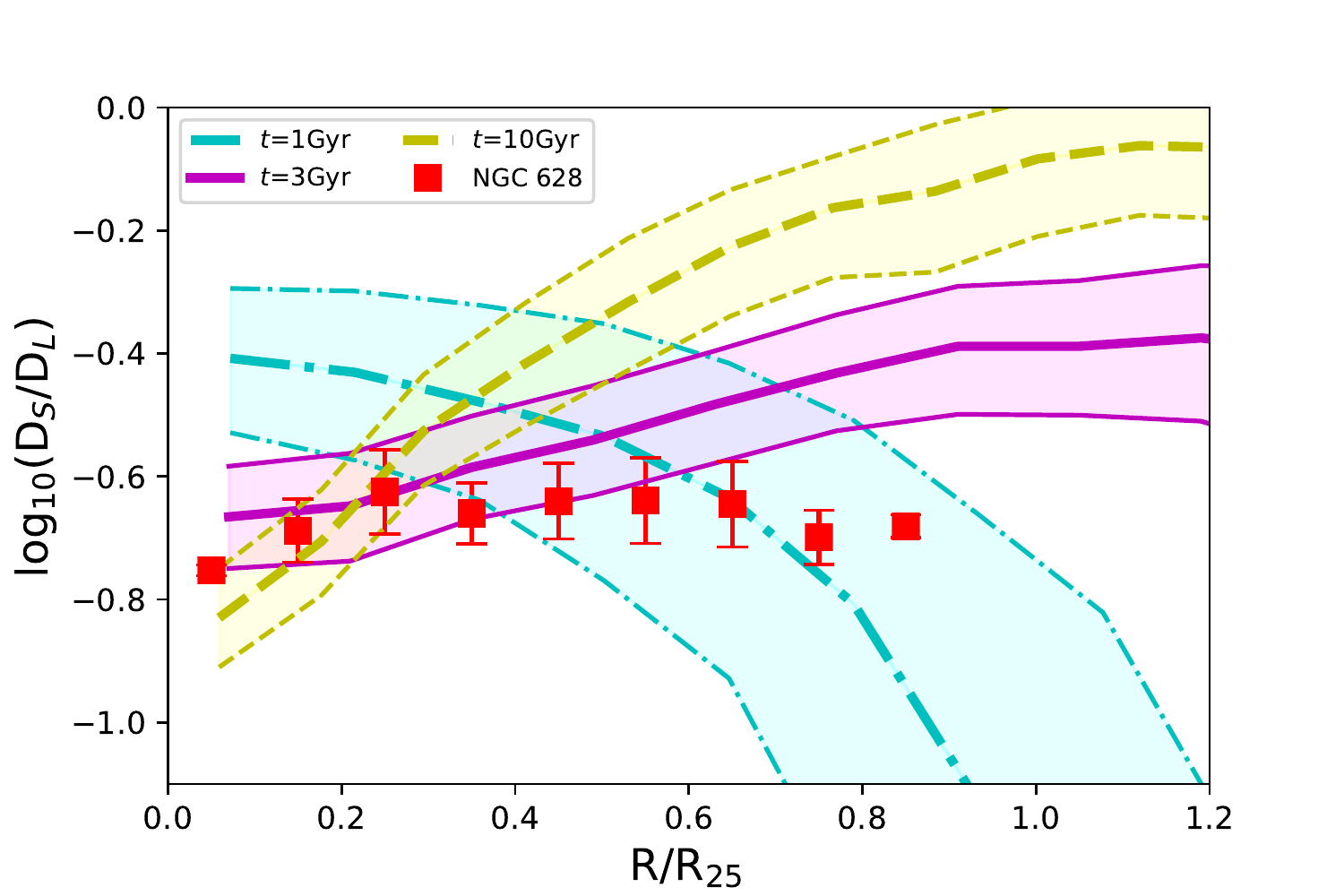}
  
 \includegraphics[width=.5\textwidth]{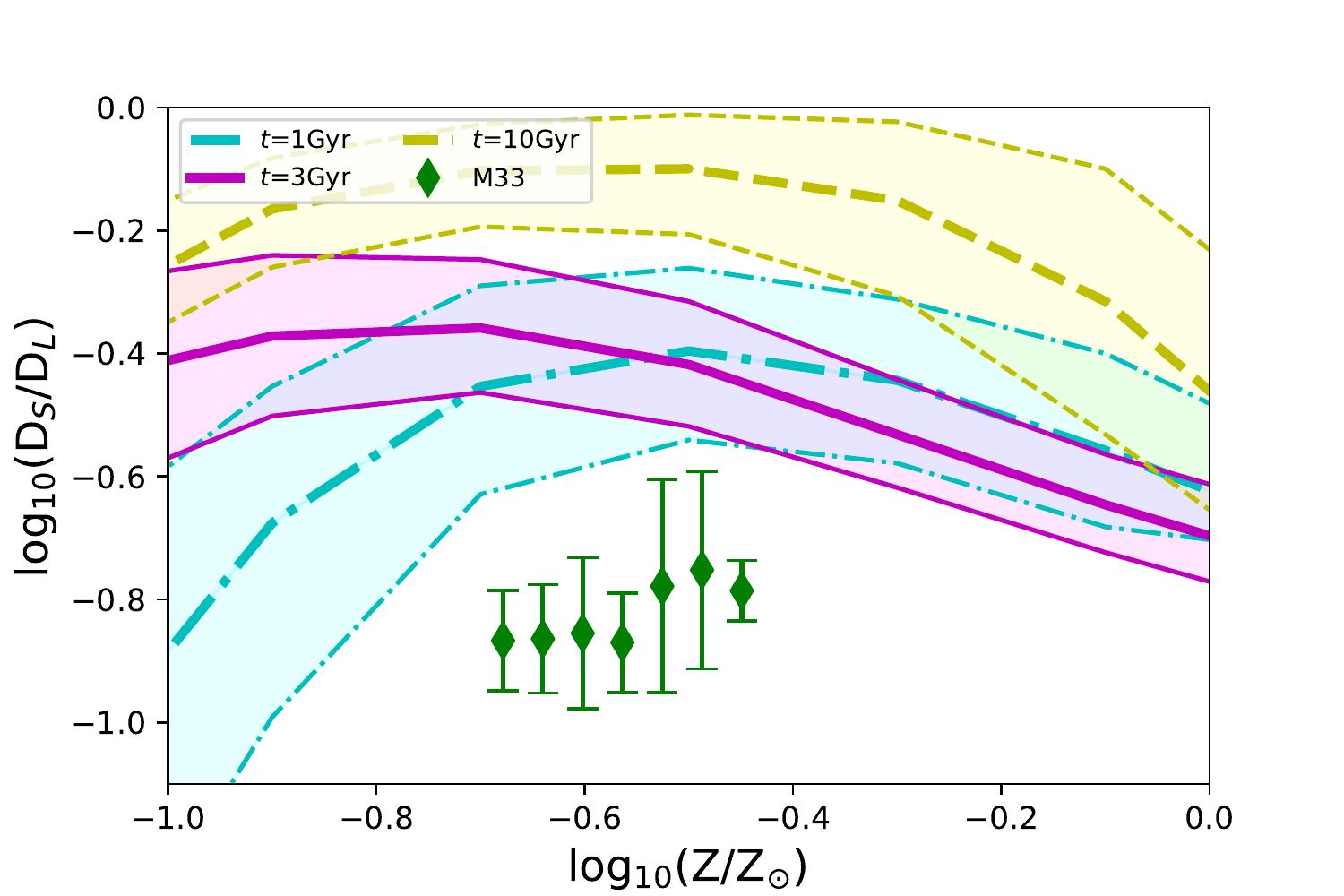}
  \includegraphics[width=.5\textwidth]{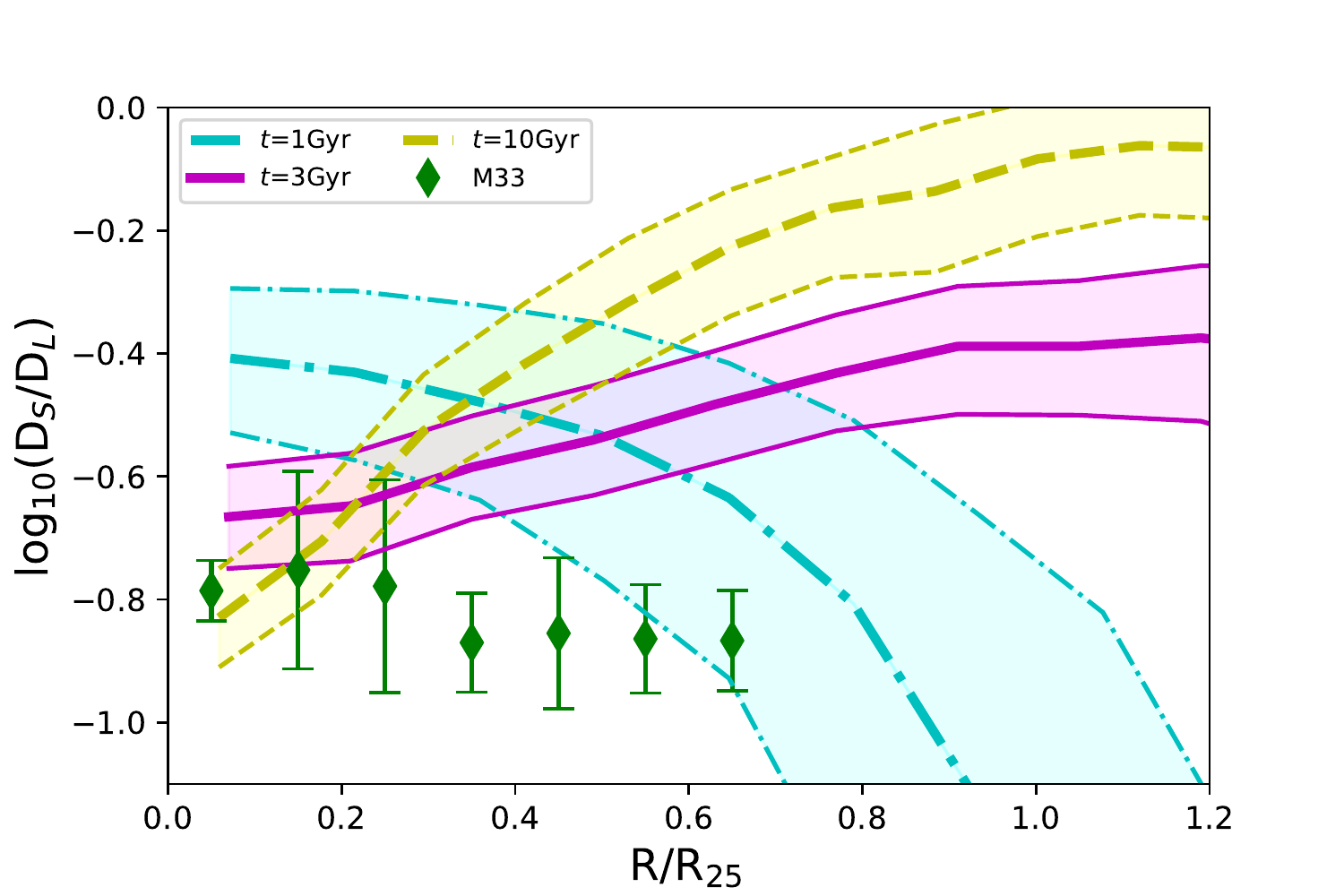}
   \caption{Comparison of the variation of the small to large grain mass ratio ($D_{S}/D_{L}$) with metallicity (left) and galactocentric radius (right) for M\,101 (top panels), NGC\,628 (middle panels), and M\,33 (bottom panels) with the SPH simulations of an isolated galaxy performed by  \citet{2017MNRAS.466..105A} and \citet{2017MNRAS.469..870H}. Thick lines represent mean values for each model and the coloured areas cover the 25 and 75 percentile of the particle distribution. A solar metallicity of Z$_{\odot}$\,=\,0.02 has been adopted.}
   \label{fig:comp_Hou17}
\end{figure*}

We present in Fig.\,\ref{fig:comp_Hou17} the ratio between the total mass of the small grains and large grains, $D_{S}/D_{L}$, as a function of metallicity (left panels) and radius (right panels). The data points were obtained integrating the $D_{S}/D_{L}$ maps derived from the SED fitting. Each point corresponds to the mean value in each ring with error bars representing the dispersion of the values within each ring. The thick lines show the predictions of the SPH simulations at different time steps of the galaxy evolution, and coloured areas show the dispersion in the models.

Within each galaxy $D_{S}/D_{L}$ seems to be independent of the metallicity and radius, especially in M\,101. However, in M\,33 we see a mild steady radial decline, though at 1$\sigma$ level, and in NGC\,628 the ratio tends to drop towards the inner high metallicity parts of the galaxy, with a central value clearly lower than the rest of the disc. Interestingly, the outer parts of M\,101 reach the same low metallicity as M\,33 but still present a constant and higher $D_{S}/D_{L}$ than M\,33. The model at 1\,Gyr reproduces the drop of the ratio at low metallicity in the outer parts of M\,33, while models of 3-10\,Gyr better describe the trend for M\,101 and NGC\,628. This is in agreement with the fact that M\,33 is expected to be a younger (and more flocculent) galaxy \citep{2017MNRAS.464.2103J} than M\,101 and NGC\,628 \citep{2013ApJ...769..127L,2014MNRAS.437.1534S}.

The radial profile of $D_{S}/D_{L}$ (right panels in Fig.\,\ref{fig:comp_Hou17}) gives hints on which dominant mechanism is influencing the evolution of the dust at each galactocentric radius. As already shown in \citet{2015MNRAS.447.2937H}, stars mainly form large grains and coagulation enhances the large grain fraction; the small grain fraction is enhanced by shattering in the diffuse medium and accretion of gas phase metals onto dust. On the other hand, shock destruction in SNe will make both large and small grain fractions to decrease. Time scales for each mechanism depend on different physical quantities: the accretion time scale is inversely proportional to metallicity and the fraction of dense gas, while the coagulation time scale depends only on the fraction of dense gas as well as on the amount of small grains. Shattering mainly occurs in diffuse areas with densities below a certain threshold \citep[$n_{\rm th}^{\rm SH}$\,=\,1\,cm$^{-3}$,][]{2017MNRAS.466..105A}. Its time scale is therefore related to the SN rate (and thus to SFR), as the shock destruction time scale.

\begin{figure*} 
 \includegraphics[width=.5\textwidth]{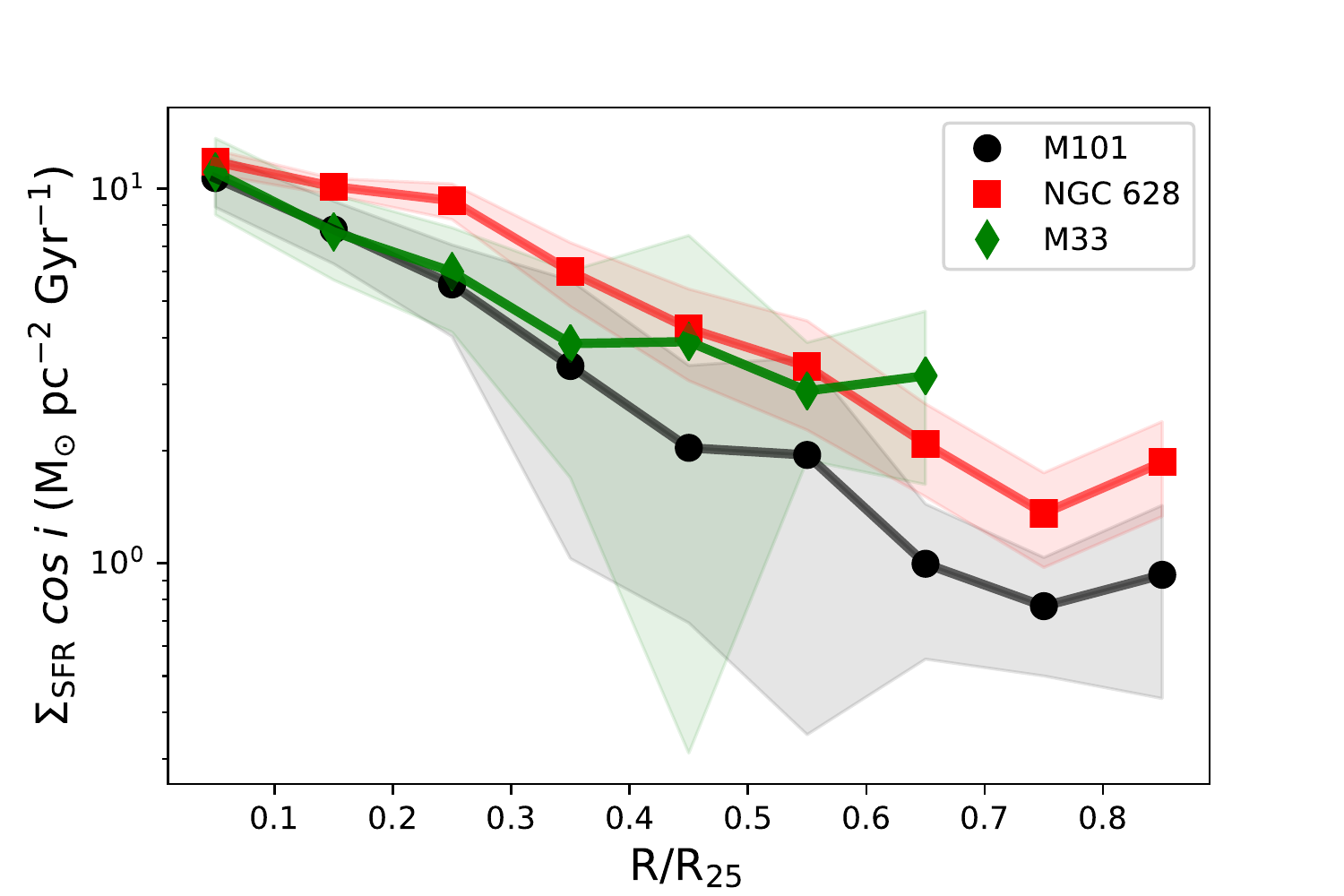}
  \includegraphics[width=.5\textwidth]{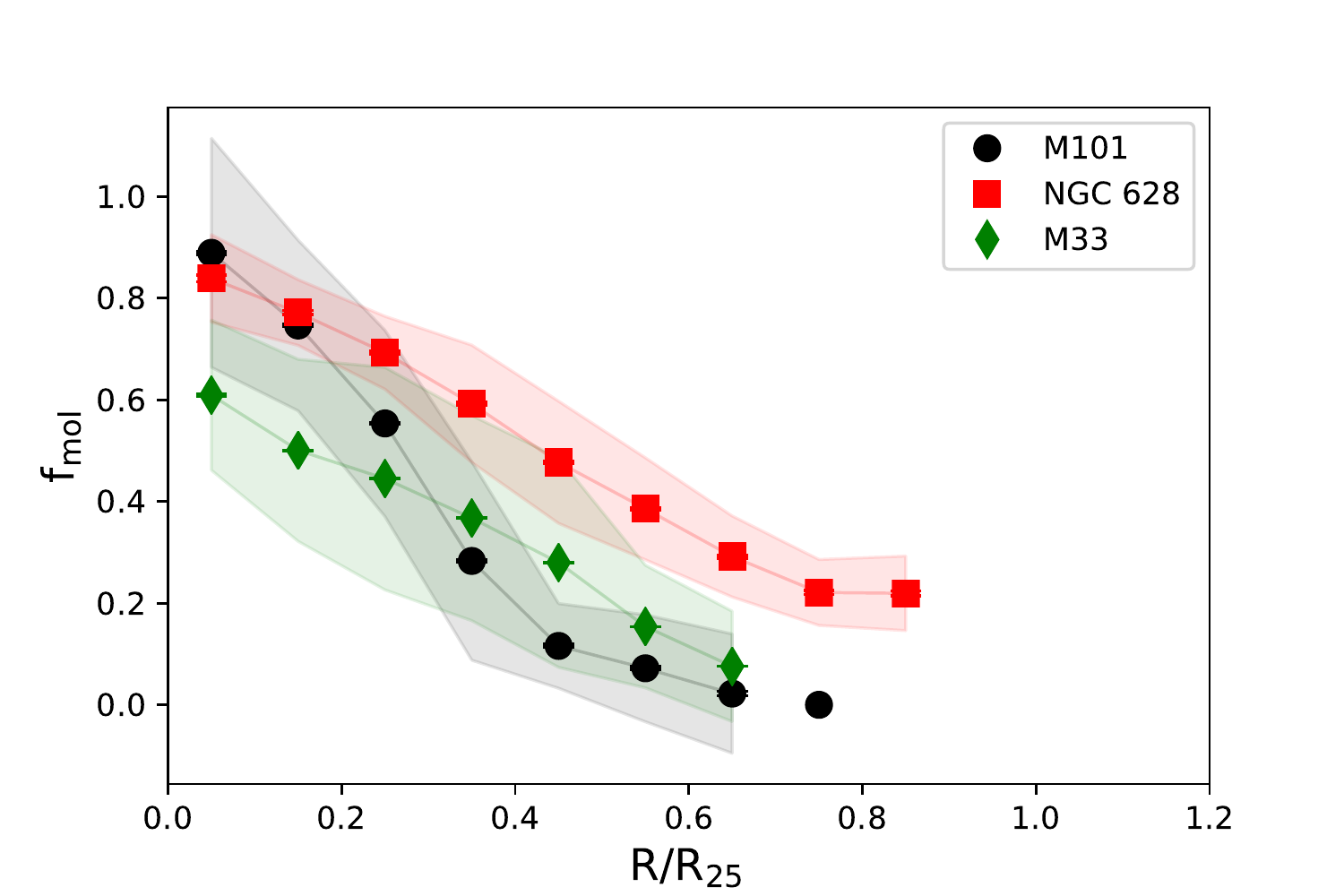}
   \caption{Radial profile of the SFR (left) and molecular gas fraction (right) (normalised to $\rm R_{\rm 25}$) for NGC\,628 (red squares), M\,101 (black circles) and M\,33 (green diamonds). SFRs were derived using FUV (GALEX) and TIR luminosities (see Eq.\,\ref{eq:sfr}). The molecular gas fraction was derived from HI and CO observations assuming a $\rm X_{CO}$ factor inversely proportional to the metallicity \citep[see][for more details]{2019MNRAS.483.4968V}. The elliptical rings have a width of 0.1\,R$_{\rm 25}$. Data points and coloured areas correspond respectively to the mean value and the dispersion for each ring.}
   \label{fig:SFR_mol}
\end{figure*}

We explain the radial behaviour of $D_{S}/D_{L}$ for our three galaxies in comparison with the results of the simulations and the radial profiles of the SFR and the molecular gas fraction for each galaxy. These are presented in Fig.\,\ref{fig:SFR_mol}. The SFR has been derived for each galaxy using FUV (GALEX) and total infrared (TIR) luminosities, and the prescription given in \citet{2011ApJ...741..124H}: 
\begin{equation}\label{eq:sfr}
\rm SFR(M_{\odot}/yr^{-1}) = 4.42\times 10^{-44}(L_{\rm FUV} + 0.49\times L_{\rm TIR})
\end{equation}
with L$_{\rm FUV}$ and L$_{\rm TIR}$ in erg\,s$^{-1}$. The TIR luminosity was derived using all the bands included in the SED fitting and the parameterisation of \citet{2011AJ....142..111B}.

The molecular gas fraction was obtained from HI and CO observations assuming a $\rm X_{CO}$ factor inversely proportional to the metallicity (see \citet{2019MNRAS.483.4968V} and \citet{2018A&A...613A..43R} for more details). We make the proxy that the molecular gas fraction describes the dense gas fraction of the galaxy, which seems plausible as the molecular gas is in general the densest gas in the ISM.

There is a drop of $D_{S}/D_{L}$ in the central parts of NGC\,628, where the metallicity and the molecular gas fraction are high (see middle panels of Fig.\,\ref{fig:SFR_mol}). In these central areas dust is growing via accretion and $D_{tot}$ is increasing (see both panels of Fig.\,\ref{fig:D2Gcomp_Hou17}) with metallicity. The {\it critical metallicity} defined by \citet{2013EP&S...65..213A} has been reached and the molecular gas fraction is quite high, which are the conditions for accretion to take place. However, in the most inner parts, at R\,$\leq$\,0.3\,R$_{\rm 25}$, coagulation is taking over and a high fraction of large grains is forming, causing $D_{S}/D_{L}$ to decrease. The coagulation time scale depends inversely on the abundance of small grains and the dense gras fraction \citep[e.g. Eq. 22 in][]{2017MNRAS.466..105A} and both quantities are high in the central parts of NGC\,628: small grains are initially abundant due to accretion and the molecular gas fraction is above f$_{\rm mol}\geq$\,0.7.

The behaviour of $D_{S}/D_{L}$ for M\,33 is completely different from NGC\,628. The ratio increases in the inner region (although at 1\,$\sigma$ level) and shows a drop in the outer R\,$\geq$\,0.3\,R$_{\rm 25}$ part of the disc. Only the model at 1\,Gyr shows an increase of $D_{S}/D_{L}$ in the central parts of the galaxy, corresponding to early phases where the high SFR produces high dust formation by stellar sources in the form of large grains but also shattering is dominant, and therefore, an effective conversion of large grains into small grains is happening. Shattering is produced by large grains colliding in shocks in the diffuse ISM. The shattering time scale is inversely dependent on the large grain abundance and the gas density. M\,33 presents a significant increase of the SFR in the central areas while the molecular gas fraction is relatively low (f$_{\rm mol}\leq$\,0.6, Fig.\,\ref{fig:SFR_mol}), which makes shattering the most favourable process in these regions.

The radial variation of $D_{S}/D_{L}$ for M\,101 is relatively constant. In the central areas (R\,$\leq$\,0.2\,R$_{\rm 25}$) the metallicity is high and the molecular gas fraction is above 0.7. Accretion is occurring in the central areas but the effect might well be balanced by coagulation, whose time scale inversely depends on the dense gas fraction, and shattering as SFR increases significantly in the central  R\,$\leq$\,0.4\,R$_{\rm 25}$ areas (left panel Fig.\ref{fig:SFR_mol}). In the outer parts of M\,101 the dense gas fraction drops significantly, the metallicity is also low and therefore, accretion and coagulation are not expected to be important. The balance between stellar dust production in the form of large grains and shattering would keep $D_{S}/D_{L}$ constant. 

\section{Comparison with cosmological simulations}\label{sec:S2L_prop}

In this section we extend the comparison of $D_{S}/D_{L}$ to a large galaxy sample. We derive a value of $D_{S}/D_{L}$ for each galaxy using the integrated SED and compare the results with the predictions of cosmological simulations. \citet{2019MNRAS.485.1727H} and \citet{2018MNRAS.478.4905A} performed SPH simulations and followed the evolution of the dust properties within a cosmological volume. They assumed the two grain size approximation from \citet{2015MNRAS.447.2937H} and dust reprocessing in the ISM: grain destruction by SN shocks, grain disruption by shattering in the diffuse ISM and grain growth by coagulation and accretion. Unlike the single galaxy simulations performed by \citet{2017MNRAS.466..105A}, the cosmological simulations made by \citet{2019MNRAS.485.1727H} and \citet{2018MNRAS.478.4905A} do include dust formation by AGB stars, in addition to dust formation by SNe. Dust grains are additionally destroyed by sputtering in hot gas ($>$10$^{6}$\,K) regions in the circumgalactic medium. While \citet{2018MNRAS.478.4905A} did not include AGN feedback and focused more in the overall dust properties of the cosmological volumen, \citet{2019MNRAS.485.1727H} included AGN feedback and studied the dust scaling relations in galaxies as well as the evolution of the grain size distribution. We compare here our results with the predictions of \citet{2019MNRAS.485.1727H}. 

\begin{figure} 
    \includegraphics[width=.5\textwidth]{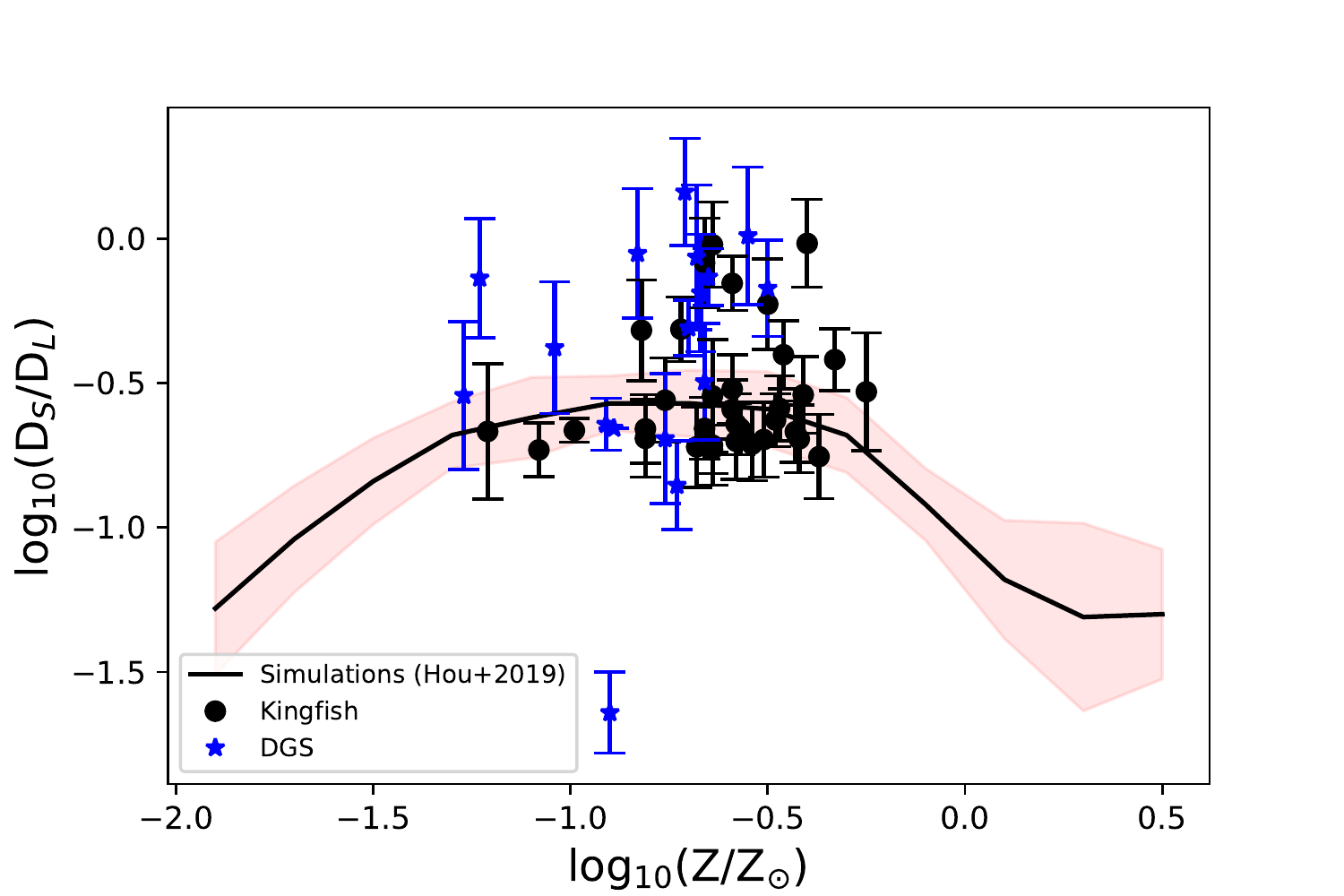}
      \includegraphics[width=.5\textwidth]{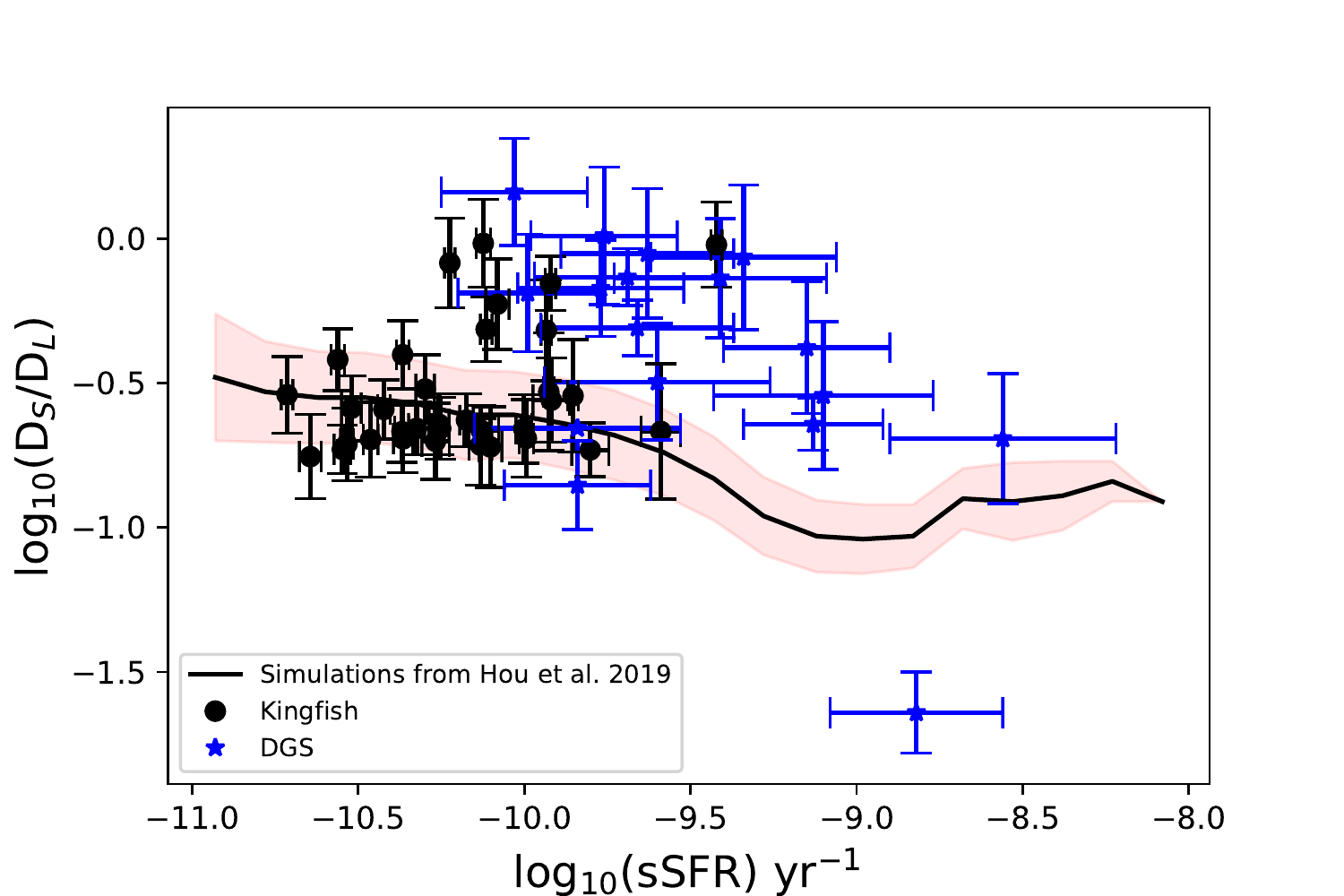}
      \includegraphics[width=.5\textwidth]{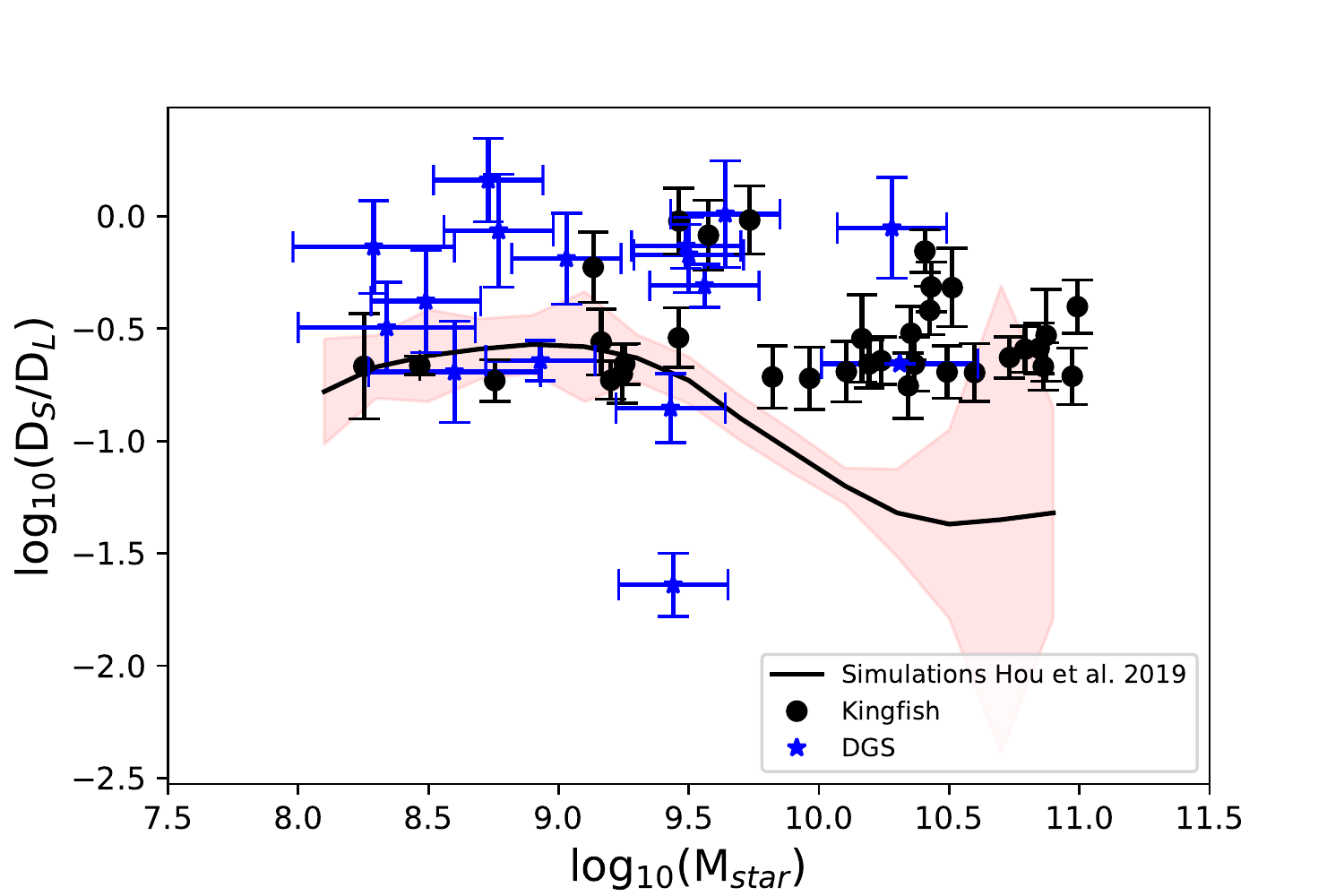}
        \caption{Comparison of $D_{S}/D_{L}$ with metallicity (top), sSFR (middle) and stellar mass  (bottom) obtained from cosmological simulations from \citet{2019MNRAS.485.1727H} and the results obtained from SED modelling for the KINGFISH (black dots) and DGS (blue asteriscs). SFRs were derived for KINGFISH galaxies using the updated FUV (GALEX) and TIR luminosities given in \citet{2017ApJ...837...90D}, while the SFR for the DGS were taken from \citet{2015A&A...582A.121R} who used  \ha\ and TIR. Stellar masses for both samples were obtained using the prescription given in \citet{2012AJ....143..139E}.
}
   \label{fig:comp_Hou19}
\end{figure}

In the top panel of  Fig.\,\ref{fig:comp_Hou19} we show the mean value of $D_{S}/D_{L}$ (black thick line) of the simulated galaxy distribution as a function of the metallicity. The coloured area represents the 1\,$\sigma$ limit of the galaxy distribution. For low metallicity galaxies, log(Z/Z$_{\odot}$)$\sim$\,-2, dust is mainly produced by stars, and shattering is the only source of small grains. At higher metallicity (-2\,$\leq$\,log(Z/Z$_{\odot}$)\,$\leq$\,-1), accretion becomes efficient and $D_{S}/D_{L}$ start to  increase significantly. At -1\,$\leq$\,log(Z/Z$_{\odot}$)\,$\leq$\,-0.5, coagulation becomes efficient and a balance between the amount of small grains created by accretion and shattering, and the large grains created via coagulation is achieved. The result is a constant $D_{S}/D_{L}$ in this metallicity range. At higher metallicities, log(Z/Z$_{\odot}$)\,$\geq$\,-0.5, coagulation dominates against accretion and shattering, and therefore the ratio decreases with increasing Z. In the bottom panel of Fig.\,\ref{fig:comp_Hou19} we show the variation of $D_{S}/D_{L}$ as a function of the stellar mass of the galaxy. The simulations present a similar behaviour as in the top panel, which reflects that the simulations reproduce in general the expected mass--metallicity correlation \citep[see Fig.\,2 in][]{2019MNRAS.485.1727H}. In the middle panel of  Fig.\,\ref{fig:comp_Hou19} a mild trend of low $D_{S}/D_{L}$ at higher specific SFR (sSFR) is observed. The trend is however modulated by the metallicity, as it can be seen in the full galaxy distribution shown in Fig. 6c of \citet{2019MNRAS.485.1727H}.

We have compared the trends predicted by the cosmological simulations in Fig.\,\ref{fig:comp_Hou19} with the $D_{S}/D_{L}$ derived from fitting the integrated SEDs of the KINGFISH and DGS galaxies.   
In general there is a relatively good agreement between $D_{S}/D_{L}$ predicted by the simulations and those derived from the SED fitting. In the top panel of Fig.\,\ref{fig:comp_Hou19} we see that, despite some KINGFISH and DGS galaxies having a high $D_{S}/D_{L}$,  most of the galaxies fall within the relation found by the cosmological simulations. 

The trend of $D_{S}/D_{L}$ versus sSFR (middle panel of Fig.\,\ref{fig:comp_Hou19}) is well reproduced in general by the KINGFISH galaxies, except for 7 galaxies having log($D_{S}/D_{L}$) above $\sim$-0.3 . Almost all the DGS shows higher $D_{S}/D_{L}$ values than the simulated ones at a given sSFR. It seems the simulations are not able to completely trace the evolution of the large and small grains for the DGS galaxies. It is possible that the dwarf galaxies, with more porous ISM and harder radiation fields than the normal spirals might process the dust grains in a different way, enhancing the mechanisms such as shattering that convert large grains into small ones \citep[e.g.][]{2017A&A...601A..55C,2018ARA&A..56..673G}. 

Finally, the simulations are not able  to trace the behaviour of $D_{S}/D_{L}$ versus stellar mass (bottom panel of Fig.\,\ref{fig:comp_Hou19}) of the KINGFISH and DGS galaxies: most of the galaxies present higher $D_{S}/D_{L}$ values than those predicted by the simulations within two orders of magnitude in stellar mass. Moreover, there is trend of constant $D_{S}/D_{L}$ at high stellar mass, which does not agree with the steep decrement of  $D_{S}/D_{L}$ predicted by the simulations at high stellar mass. The steep decrement of $D_{S}/D_{L}$ could be related to the AGN feedback model included in the simulations. AGN feedback is added to suppress the metal enrichment and explain the deficit of galaxies in the high mass end of the stellar mass function \citep[see Fig.\,1 in][]{2019MNRAS.485.1727H}. As the accretion time scale depends on the inverse of the metallicity, a decrement of metallicity in massive galaxies would lead to less accretion, and therefore less dust mass in the form of small grains. Besides, if these simulated galaxies still have a high dense gas fraction, coagulation can also be important and more dust in the form of large grains would be expected. Since the simulations overpredict the dust-to-gas mass ratio in massive galaxies  \citep[see Fig.\,4b in][]{2019MNRAS.485.1727H}, increasing the accretion efficiency would move the dust-to-gas mass ratio further away from the observations, and therefore a higher dense gas fraction in the simulated galaxies would be probably the best possible explanation.

The trend of $D_{S}/D_{L}$  as a function of metallicity, sSFR and stellar mass predicted by the simulations and derived from the observations is, in general, in reasonable agreement considering that we are comparing galaxies with different morphological types and at different evolutionary states and physical conditions of the ISM. This comparison reinforces the general trends and results of the cosmological simulations in tracking the evolution of the dust grain size distribution and gives some clues about possible limitations and caveats in the assumptions and initial conditions adopted in the simulations. 

\section{Discussion}\label{sec:disc}
The work presented here shows a set of observational results that can be used to test the validity of the assumptions of the simulations and models. We find agreements and disagreements between observations and simulations. In some cases the disagreements are due to a poor match of the observational sample with the physical conditions assumed in simulations, and in others the comparison might reveal problems in the simulations themselves. In the later case, the comparison to the observations might serve as a guide for further improvements in future simulations. In this section we discuss in more detail these agreements and disagreements.

\subsection{Disagreement between observations and simulations}

The main disagreements in our comparison between observations and simulations for the spatially resolved studies are: i) flatter observational dust radial profiles than those predicted by the simulations, and ii) higher $D_{tot}$ derived from the observations at low metallicities. For integrated galaxies, the behaviour of the observed $D_{S}/D_{L}$  as a function of sSFR and stellar mass is not well reproduced by the simulations. 

Some of the discrepancies are due to physical reasons because our observed galaxies do not resemble completely the galaxy types assumed in the simulations. For the case of the spatially resolved studies, the simulations predict higher central dust surface densities and steeper radial gradients than those provided by the observations. This reflects the difference of the initial conditions adopted by the single galaxy SPH simulations, which assume a galaxy with a large bulge. None of our galaxies shows evidence of large bulges in their centres: our set of galaxies are late-types rather than earlier prominent bulge types assumed to describe the initial conditions in the SPH simulations. In a similar way, when comparing the integrated SEDs of the DGS galaxies to cosmological simulations, we find that the physical properties of the ISM for low-metallicity galaxies of the DGS is not well reproduced by the simulations, which are more adequate to describe higher metallicity galaxies. In order to properly describe the DGS galaxies, a harder ISRF and a more porous ISM would have to be included in the simulations. This would most likely make the comparison of $D_{S}/D_{L}$ versus sSFR and metallicity to agree better with the observations.

Other discrepancies shown in this study are more related to limitations of the assumptions in the simulations. For spatially resolved studies, the discrepancy in the dust mass surface density in the central parts of the galaxy and the agreement in the $D_{tot}$ in the same areas indicate that the gas mass fraction assumed in the simulations might be too high. This discrepancy has been already suggested by previous studies  \citep{2017ApJS..233...22S,2015MNRAS.447.1610M}. Besides, observed radial gradients of $D_{tot}$ agree better with simulations than the observed behaviour of $D_{tot}$ as a function of metallicity, which shows that simulations predict larger azimuthal variations in metallicity than what it is typically observed. Finally, AGB stars are not taken into account within the frame work of the SPH simulations of a single galaxy of \citet{2017MNRAS.466..105A} and \citet{2017MNRAS.469..870H}. If the expected contribution of the AGB stars to the dust budget is higher at low metallicities, as it has been suggested previously \citep[e.g.][]{2009MNRAS.397.1661V}, this would affect the outer parts of the galaxies and might make the gradient of $D_{tot}$ versus metallicity to agree better with the observations. 

For the case of integrated galaxies, the observational behaviour of the $D_{S}/D_{L}$ ratio as a function of the stellar mass is not 
well reproduced by the cosmological simulations, especially in the higher stellar mass range. The discrepancy might be related to the prescription for AGN feedback assumed in the cosmological simulations, which is able to reproduce the deficit of galaxies in the high mass end of the stellar mass function but predicts lower metallicities. Environments with low metallicity would lead to less accretion and less dust mass in the form of small grains. Therefore the prescription for AGN feedback underpredicts the $D_{S}/D_{L}$ ratio for the  most massive galaxies.

\subsection{Agreements and conclusions on dust evolution}

For spatially resolved observations there is a general agreement of the radial relations of $D_{tot}$ and $D_{S}/D_{L}$ ratio between observations and simulations. The spatially resolved studies of the three different galaxies reveals that the evolution of the dust is not well described by one single parameter. Coagulation, accretion and shattering affect the $D_{S}/D_{L}$ ratio, but the time scales of these processes also depend on other physical parameters such as metallicity, SFR and  f$_{\rm mol}$. Therefore, the $D_{S}/D_{L}$ ratio, in combination with the SFR and f$_{\rm mol}$, can be used to indicate the main mechanism that is responsible for the evolution of the interstellar dust. 
Based on this scenario we can interpret radial profiles of these parameters for our three galaxies and derive which is the most important mechanism in the dust evolution at different galactocentric distances. While in the most central regions of NGC~628 there is evidence that coagulation is taking over other mechanisms, in M\,101 accretion seems to be in equilibrium with coagulation and shattering. This is not the case of M\,33, where in the central areas shattering is the most favourable mechanism regulating the evolution of the interstellar dust. 

The integrated values of $D_{S}/D_{L}$ for the galaxy sample considered in this study show a general agreement with the predictions of the variation of this ratio with metallicity. Most of the observational ratios from the KINGFISH galaxy sample lie within the limits predicted by the simulations and within a metallicity range where a combination of accretion, coagulation and shattering is driving the evolution of the interstellar dust confirming that the simulations describe the dust evolution in this type of galaxies well.

\section{Conclusions}\label{sec:con}

We study in this paper how the total mass of the small and large grains varies across the disc of a set of spiral galaxies with different masses. This observational information is needed to compare and validate the recent results of the SPH simulations that include evolution of the grain size distribution. We also study the variation of $D_{S}/D_{L}$ ratio for a large galaxy sample and compare it with the predictions of the recent cosmological simulations.    

Using the results provided by fitting the observed SED at each location of the disc of a set of three spiral galaxies: NGC\,628, M\,101 and M\,33, we are able to derive the relative abundance of each grain type across the individual galactic discs. We have applied the same fitting procedure to the KINGFISH and DGS galaxies and compare the $D_{S}/D_{L}$ ratio derived from the fitting with the predictions of the cosmological simulations. Our main conclusions are the following: 
\begin{itemize}
\item The radial profile of the dust mass surface density provided by the simulations of a single galaxy shows higher values than the observations in the centres of the galaxies. The simulated dust mass distribution at ages of 1 to 10\,Gyr agree reasonably well at larger galactocentric distances. The differences in the central parts might be related to the initial conditions of the galactic disc assumed by the simulations. 
\item  The agreement between $D_{tot}$ derived from observations and the predictions of the simulations is good at high metallicities and in the central parts of the galaxies. For low metallicities the agreement is worse, which most likely is due to the fact that the metallicity variations predicted by the simulations are large and affected by the initial conditions assumed by the simulations.
\item The behaviour of the $D_{S}/D_{L}$ ratio as a function of radius and metallicity shows different trends for each galaxy, revealing the importance of the main mechanism regulating dust evolution. For NGC\,628 the ratio declines in the central parts, which shows that coagulation is important in these areas. For M\,33 the ratio increases towards the central parts presenting evidence that shattering is the main mechanism affecting the dust evolution. For M\,101 the ratio is constant across the whole disc, which suggests that in this galaxy there is a balance between accretion and shattering forming small grains, and coagulation and star formation as main producers of large grains. 
\item We compare the $D_{S}/D_{L}$ ratio derived from the integrated SEDs of KINGFISH and DGS galaxies with the  results of the cosmological simulations performed by \citet{2019MNRAS.485.1727H}. In general the agreement between predictions from the simulations and results from the SED fitting is quite reasonable. Simulations seem to explain well the behaviour of $D_{S}/D_{L}$ as a function of metallicity. The behaviour of $D_{S}/D_{L}$ versus stellar mass for both galaxy samples is not well reproduced by the simulations, especially in the high stellar mass end.  
\end{itemize}

We present here, for the first time, a comparison of the dust-to-gas mass ratio and small to large grain mass ratio predicted by SPH simulations of a single galaxy including evolution of the grain size distribution with the results provided by fitting the observational SEDs. The comparison gives an observational support to the results provided by the simulations. A detailed simulation of each individual galaxy is out of the scope of the paper. Cosmological simulations including evolution of the grain size distribution find here a set of observational data which can be useful in order to test their results and improve their assumptions.

\begin{acknowledgements}
The authors would like to thank the referee, Prof. Takeuchi, for very constructive comments that have helped to improve first version of the paper. MR and UL acknowledges support by the research projects AYA2014-53506-P and AYA2017-84897-P from the Spanish Ministerio de Econom\'{\i}a y Competitividad and Junta de Andaluc\'{\i}a grant FQM108; support from the European Regional Development Funds (FEDER) is acknowledged. KCH is supported by the IAEC-UPBC joint research foundation (grant No. 257) and Israel Science Foundation (grant No. 1769/15). MR and UL thank to the Computational service PROTEUS at the Instituto Carlos I. This work was partially supported by the Spanish Ministerio de Econom\'{\i}a y Competitividad under grants AYA2016-79724-C4-4-P and AYA2016-79724-C4-3-P, and excellence project PEX2011-FQM-7058 of Junta de Andaluc\'{\i}a (Spain). This research made use of APLpy, an open-source plotting package for Python hosted at http://aplpy.github.com of TOPCAT \& STIL: Starlink Table/VOTable Processing Software of Matplotlib, a suite of open-source python modules that provide a framework for creating scientific plots.

 \end{acknowledgements}

\bibliographystyle{aa} 
\bibliography{references}  
\end{document}